\documentclass[prd, 12pt,a4paper,groupedaddress,preprintnumbers,nofootinbib,floatfix]{revtex4-1}
\usepackage[top=2.9cm, bottom=2.1cm, left=2cm, right=2cm]{geometry}
\usepackage[english]{babel}
\usepackage[utf8]{inputenc}
\usepackage{amsmath,amssymb,amsfonts, dsfont}
\usepackage{graphicx}
\usepackage{color}
\usepackage{hyperref}
\usepackage{subfigure}
\usepackage{dcolumn}
\usepackage{bm}
\usepackage{float}
\usepackage{mathtools}
\usepackage{amsthm}
\usepackage{bbm}
\usepackage{pxfonts}
\usepackage[vcentermath]{youngtab}
\usepackage[all]{xy}
\usepackage{pstricks}
\usepackage{accents}
\usepackage{cases}
\usepackage{comment}
\usepackage{placeins}
\usepackage{xspace}
\usepackage{cancel} 
\usepackage{slashed}
\usepackage{soul}
\usepackage{feynmp}
\usepackage{ytableau}
\usepackage{youngtab}
\usepackage{multirow}
\usepackage{feynmp-auto} 

\definecolor{orange}{cmyk}{0,0.5,1,0}
\definecolor{rossoCP3}{cmyk}{0,.88,.77,.40}
\definecolor{graa}{rgb}{0.8,0.8,0.8}
\definecolor{blaa}{rgb}{0.2,0.2,0.6}
\hypersetup{
	colorlinks, 
	bookmarksopen, 
	bookmarksnumbered,
	citecolor=blaa, 		
	linkcolor=rossoCP3,	
	urlcolor=rossoCP3,			
	    }

\newcommand{\cL}{\mathcal{L}}

\newcommand{\f}{\mathcal{F}}
\newcommand{\s}{\mathcal{S}}

\newcommand{\be}{\begin{equation}}
\newcommand{\ee}{\end{equation}}

\newcommand{\TC}{\mathrm{TC}}
\newcommand{\SM}{\mathrm{SM}}
\newcommand{\SU}{\mathrm{SU}}
\newcommand{\Sp}{\mathrm{Sp}}

\newcommand{\UU}{\mathrm{U}}
\newcommand{\LL}{\mathrm{L}}
\newcommand{\RR}{\mathrm{R}}

\newcommand{\tcf}{\mathcal{F}}
\newcommand{\tcs}{\mathcal{S}}
\newcommand{\transpose}{^{\mathrm{T}}}
\newcommand{\hc}{\; + \; \mathrm{h.c.} \;}
\newcommand{\Tr}{\text{Tr}}
\newcommand{\andeq}{\quad \mathrm{and} \quad}

\newcommand{\brakets}[1]{\left\langle #1 \right\rangle}
\newcommand{\abs}[1]{\left| #1 \right|}

\newcommand{\spur}[2]{\psi^{#1}\phantom{}_{#2} }
\newcommand{\spurbar}[2]{\bar{\psi}^{#1 #2}} 

\newcommand{\Yf}[4]{( y_f^\ast y_f)^{a_{#1}}\phantom{}_{a_{#2}}\phantom{}^{i_{#3}i_{#4}}}
\newcommand{\SDS}[2]{(\Sigma^\dagger \overleftrightarrow{D}^\mu\Sigma)_{a_{#1}}\phantom{}^{a_{#2}}}
\newcommand{\DS}[2]{(D^\mu\Sigma)^{a_{#1}a_{#2}}}
\newcommand{\DSd}[2]{(D_\mu\Sigma^\dagger)_{a_{#1}a_{#2}}}

\usepackage{xcolor}
\usepackage{soul}

\usepackage[noindentafter]{titlesec}
\begin{document}

 
\title{\texorpdfstring{\Large\color{rossoCP3}   Minimal Fundamental Partial Compositeness }{}}
\author{Giacomo {\sc Cacciapaglia} ${}^a$}
\email{g.cacciapaglia@ipnl.in2p3.fr} 
\author{Helene {\sc Gertov} ${}^b$}
\email{gertov@cp3.sdu.dk} 
\author{Francesco {\sc Sannino} ${}^{b,c,d}$}
\email{sannino@cp3.dias.sdu.dk}
\author{Anders Eller {\sc Thomsen} ${}^b$}
\email{aethomsen@cp3.sdu.dk} 
\affiliation{${}^a$ Univ Lyon, Universit\'e Lyon 1, CNRS/IN2P3, IPNL, F-69622, Villeurbanne, France}
\affiliation{${}^b$CP$^3$-Origins, University of Southern Denmark, Campusvej 55, 5230 Odense, Denmark}
\affiliation{${}^c$Danish IAS, University of Southern Denmark,  Odense, Denmark}
\affiliation{${}^d$Theoretical Physics Department, CERN, Geneva, Switzerland.} 

\begin{abstract}
 Building upon the fundamental partial compositeness framework we provide consistent and complete composite extensions of the standard model. These are used to  determine the effective operators emerging at the electroweak scale in terms of the standard model fields upon consistently integrating out the heavy composite dynamics.  We exhibit the first effective field theories matching  these complete composite theories of flavour and analyse their  physical consequences for the third generation quarks. Relations with other approaches, ranging from effective analyses for partial compositeness to extra dimensions as well as purely fermionic extensions, are briefly discussed. Our methodology is applicable to any composite theory of dynamical electroweak symmetry breaking featuring a complete theory of flavour.   
\\[.3cm]
{\footnotesize  \it Preprint: CP$^3$-Origins-2017-016 DNRF90 \&  LYCEN-2017-04 \& CERN-TH-2017-093}
\end{abstract}
\maketitle
\newpage

\tableofcontents

\section{Introducing the Minimal Fundamental Composite Model} 

Since the earliest proposals of new composite dynamics (aka Technicolour -- TC) as the underlying theory of electroweak symmetry breaking~\cite{Weinberg:1975gm,Dimopoulos:1979es}, generating masses for the Standard Model (SM) fermions has been the biggest hurdle on the way to a complete model. Many attempts have been made, from extending the TC gauge sector~\cite{Eichten:1979ah} to introducing scalar mediators as in Bosonic TC~\cite{Simmons:1988fu,Samuel:1990dq,Kagan:1991gh,Carone:1992rh,Carone:1994mx,Antola:2009wq}.
The SM fermion masses are generated either by effective operators bilinear in the fermion spinors, or via linear mixing to a fermionic bound state as in the partial compositeness mechanism~\cite{Kaplan:1991dc}. In all cases, the main difficulty has been to construct a complete theory in the ultra-violet. 
Phenomenologically it is difficult to accommodate a heavy top quark with the stringent bounds on the scale of flavour violation in the light quark and lepton sectors.
Recently, in Ref.~\cite{Sannino:2016sfx}, an alternative paradigm has been introduced that allows for writing a complete UV theory of composite flavour. The models  simultaneously account for a pseudo Nambu Goldstone Boson (pNGB) Higgs \cite{Kaplan:1983fs}, and can be extrapolated to the strong gravity scale.  Here fermion masses are generated via Yukawa couplings involving TC-charged scalars. Partial compositeness is thus obtained at low energy by the formation of fermion-scalar bound states. Composite theories including (super) TC scalars, attempting to give masses to some of the SM fermions, appeared earlier in the literature \cite{Kagan:1991ng,Dobrescu:1995gz,Kagan:1994qg,Antola:2010nt,Antola:2011bx,Altmannshofer:2015esa}  for (walking) TC theories that did not feature a pseudo Nambu Goldstone Boson Higgs. 

In models of Fundamental Partial Compositeness (FPC) the SM is extended with a new TC sector featuring new elementary fermions and scalars charged under a new gauge group $ G_\TC $~\cite{Sannino:2016sfx}.  Electroweak symmetry breaking (EWSB) is caused by the TC dynamics in which the Higgs boson is replaced by a light composite state.

The TC Lagrangian before introducing the electroweak sector reads:  
\begin{equation}
\mathcal{L}_{\mathrm{TC}} = -\tfrac{1}{4} \mathcal{G}_{\mu\nu} \mathcal{G}^{\mu\nu} + i \overline{\tcf} \bar{\sigma}^\mu D_\mu \tcf - \left(\tfrac{1}{2} \tcf m_\tcf \epsilon_\TC \tcf \hc\right) + \left(D_\mu \tcs\right)^{\dagger} \left(D^{\mu} \tcs\right) -  \tcs^{\dagger} m^2_\tcs \tcs - V(\tcs),
\label{eq:L_kin}
\end{equation} 
where  TC-fermions and TC-scalars are in pseudoreal representations of the  $ G_\TC $ group, 
 $ m_\tcf $ and $ m^2_\tcs $ are mass matrices and $ \epsilon_\TC $ is the antisymmetric invariant tensor of $ G_\TC $.  This choice of representation is due to the fact that the most minimal models are of this nature \cite{Cacciapaglia:2014uja}. Nevertheless the following analysis and methodology is generalisable to complex and real representations as well, for which a list of FPC models was made in \cite{Sannino:2016sfx}.  
 
Assuming $N_\tcf$ Weyl TC-fermions the maximal quantum global symmetry of the fermions in the kinetic term is $\SU(N_\tcf)$. The symmetries are such that $ m_\tcf$ is an antisymmetric tensor in flavour space. 

As the TC-scalars transform according to the same representation as the TC-fermions with respect to the new gauge group, no Yukawa interactions among the TC-fermions and TC-scalars can be written (except for few exceptions~\cite{Sannino:2016sfx}). This implies that, with zero mass terms, the TC-scalars have an independent $\Sp(2N_\tcs)$ symmetry. We assume the potential $V(\tcs)$ to respect the maximum global symmetries of the TC theory.  To elucidate the symmetry in the scalar sector we note that the $N_\tcs$ complex TC-scalars can be arranged in the following single field
	\begin{equation}
	\Phi = \begin{pmatrix} \tcs \\ -\epsilon_\TC \tcs^{\ast} \end{pmatrix} \ ,\label{eq:Phi}
	\end{equation}
still transforming according to a pseudoreal representation of $ G_\TC $. The TC indices are hidden to keep the notation light, cf. Appendix \ref{app:definitions}.  One can show that  this rearrangement leaves the TC Lagrangian invariant under the $ \Sp(2N_\tcs) $ flavor symmetry.  The scalar kinetic and mass term now reads: 
	\begin{equation}
	\frac{1}{2} \left(D_\mu \Phi\right) \epsilon_\TC \epsilon \left(D^{\mu} \Phi\right) -  \frac{1}{2}\Phi   \epsilon_\TC M^2_\tcs \Phi \ , \label{eq:lphi}
	\end{equation}
with 
	\begin{equation} 
	M^2_\tcs   = \begin{pmatrix} 0 & -{m^2_\tcs}^T  \\ {m^2_\tcs} & 0 \end{pmatrix} \ ,
	\end{equation}
and $ \epsilon $ is the invariant symplectic form of $ \Sp(2N_\tcs) $. 
 
A straightforward realisation for this model is obtained choosing $ G_\TC = \Sp(2N) $ with the TC fundamental states in the fundamental representation.  In Table~\ref{Table-Fields} we summarise the elementary states of the TC theory as well as the bilinear gauge singlets along with their global transformation properties and multiplicities.
\begin{table}[t]
	\begin{tabular}{*{4}{c}}
	 States \qquad&  $ \SU(N_\tcf) $  \qquad&  $ \Sp(2N_\tcs) $ \qquad & number of states \qquad \\ \hline 
	$ \tcf $ & $\tiny\yng(1)$ & $ 1 $ & $2 N \times N_\tcf$  \\
	$ \Phi $ &  1 & $ {\tiny \yng(1)} $ & $2N \times 2N_\tcs$  \\
	\hline
	$ \Phi\Phi $ &  1 & $ 1 + {\tiny \yng(1,1)} $ & $ \displaystyle{1 + {N_\tcs(2N_\tcs-1)}}$ \\
	$ \tcf \Phi $ & $ {\tiny\yng(1)} $ & $ {\tiny\yng(1)} $ & $ 2N_\tcs N_\tcf $\\
	$ \tcf \tcf $ & $ {\tiny\yng(1,1)} $ & 1  & $ \displaystyle{{N_\tcf(N_\tcf-1)}}$ 
	\end{tabular}
	\label{Table-Fields}
	\caption{\emph{The fundamental matter fields of the theory appear in the first two lines of the table, both transforming according to the fundamental representation of TC.  The last three lines correspond to  the bi-linear composite TC singlet states. The number of states counts the Weyl fermions or real scalars.}}
\end{table}

When adding the electroweak (EW) sector we  embed it within the $ \SU(N_\tcf) $ of the TC-fermion sector. In this way the EWSB is tied to the breaking of $ \SU(N_\tcf) $ and the Higgs boson can be identified with a pNGB of the theory \cite{Kaplan:1983fs,Cacciapaglia:2014uja}.  Assuming for the scalars a positive mass squared, it is natural to expect spontaneous symmetry breaking in the fermion sector~\footnote{In \cite{Sannino:2016sfx} there is also a preliminary analysis of the potential conformal window including light TC-scalars that allows to argue that the model is expected to be in a chirally broken phase.} according to the pattern $\SU(N_\tcf) \rightarrow \Sp(N_\tcf)$. This breaking pattern was established in the absence of scalars for $N_\tcf = 4$ and $G_\TC=\Sp(2)$ via first principle lattice simulations~\cite{Lewis:2011zb}. The ensuing TC-fermion bilinear condensate is 
	\begin{equation}
	\brakets{\tcf^{a} \epsilon_\TC \tcf^{a'} } = f^2_\TC \Lambda_\TC \Sigma_0^{a a'},  
	\end{equation}
where Lorentz and TC indices are opportunely contracted, and the $ \Sigma_0 $ matrix is an antisymmetric, two-index representation of $\SU(N_\tcf)$. We also have $\Lambda_\TC = 4\pi f_\TC$ with $\Lambda_\TC$ the composite scale of the theory and $f_\TC$ the associated pion decay constant.
 
In addition, we envision two possibilities for the TC-scalars: the formation of a condensate $\brakets{\Phi^i \epsilon_\TC \Phi^j}$ may not happen or be proportional to the singlet of $\Sp(2 N_\tcs)$, in which case the flavour symmetry  in the scalar sector is left unbroken; or a condensate forms and breaks $\Sp(2 N_\tcs)$ generating light bosonic degrees of freedom. For the remainder of this paper we will focus on the former case for the sake of simplicity.

We now turn our attention to the SM fermion mass generation. The presence of TC-scalars in FPC models allow for a new type of Yukawa interactions interfacing the TC and the SM sectors. In fact each new Yukawa operator involves a TC-fermion, a TC-scalar and a SM fermion and the new fundamental Yukawa Lagrangian to replace the SM one reads
	\begin{equation}
	\mathcal{L}_{\mathrm{yuk}} = - \spur{i}{a} \epsilon_{ij} \Phi^{j} \epsilon_\TC \tcf^{a} \hc \ ,
	\label{eq:fund_Yukawa}
	\end{equation}  
in which we make use of the  spurion $ \psi $  transforming under the relevant global symmetries as
	\begin{equation} \label{eq:spurionpsi}
	\spur{i}{a} \equiv \left(\Psi \, y\right)^{i} \phantom{}_{a} \in {\tiny\yng(1)}_{\tcs} \otimes \overline{\tiny\yng(1)}_{\tcf} \ .
	\end{equation} 
Here $\Psi$ is a generic SM fermion and $y $ is the new Yukawa matrix. With this spurionic construction we may formally consider $ \mathcal{L}_{\mathrm{yuk}} $ an invariant of the global TC symmetries. Additionally, the notation has the benefit that all Yukawa interactions are summarised in a single operator. Note that with the notation introduced here, the generation, colour, and electroweak indices are all embedded in the global symmetries. 
At low energy, the Yukawa couplings in eq.~\eqref{eq:fund_Yukawa} generate linear mixing of the SM fermions with spin-1/2 resonances made of one TC-fermion and one TC-scalar (see Table~\ref{Table-Fields}), thus implementing partial compositeness. This way of endowing masses for the SM fermions is free from long standing problems in models of composite Higgs dynamics and, as we shall discuss later, can be also related to previous incomplete extensions. 

Besides the SM fermions and Yukawas, the underlying theory contains two more spurions that explicitly break the flavour symmetries, that is the masses of the TC-fermions and scalars:
	\begin{equation}
	m_\tcf \in  \overline{\tiny\yng(1,1)}_{\tcf} \otimes 1_\tcs\ , \quad M^2_\tcs \in 1_\tcf \otimes \overline{\tiny\yng(1,1)}_{\tcs}\ .
	\end{equation}
As they are dimension-full parameters, they can be inserted at the effective Lagrangian level only if an order parameter can be defined, i.e. either if the mass is small compared to the TC scale $\Lambda_\TC$, or if they are much larger. In the latter case, one can then expand in powers of the inverse of the mass matrices.
We will start with the former case, and classify the relevant operators in terms of powers of the spurion $ \spur{i}{a} $, and then discuss  how to consistently move to the limit of large TC-scalar masses.  

We are now ready to determine the effective operators emerging at the EW scale in terms of the SM fields upon consistently integrating out the heavy TC dynamics aside from the pNGB excitations. De facto we provide the first effective field theory that matches to a concrete  and complete example of a composite theory of flavour. In turn, this allows for investigating its impact on electroweak observables and low energy flavour physics.

We structure the work as follows. In Section \ref{EFT} we construct the effective field theory. We set the stage by first briefly reviewing the essentials of the TC pNGB effective field theory. We then move on to construct the symmetry allowed TC-induced effective  operators involving SM fermions. We construct both fermion bilinears, and four-fermions operators. Then we formulate the standard model induced one-loop pNGB potential and higher derivatives pNGB operators. Physical consequences and phenomenological constraints deriving from the third generation quarks physics are investigated in Section \ref{topandbottom}, in which we also briefly comment on the light generations. Section \ref{Connection} is devoted to the relation with other approaches ranging from effective analyses for partial compositeness to extra dimensions as well as purely fermionic extensions. We finally offer our conclusions in Section \ref{Concluisons}.

\section{Effective Field Theory at the electroweak scale}
\label{EFT}

Having spelled out the underlying fundamental dynamics we now move to determine the effective operators at the EW scale. We start with a brief summary of the chiral Lagrangian for the TC sector.  We then list the effective operators in terms of the SM fields generated by explicit realisations of partial compositeness. This is achieved by coherently matching the operators to the underlying composite flavour dynamics. This allows, for the first time, to build in a controlled manner the full effective field theory.  
 All operators will then appear in the Lagrangian 
	\begin{equation}
	\mathcal{L}_{\mathrm{EFT}} = \sum_{A} C_A \, \mathcal{O}_A + \left(\sum_{A}C'_A\, \mathcal{O}'_A \hc \right)
	\end{equation}
for the effective field theory with coefficients $ C^{(\prime)}_A $ determined by the underlying TC dynamics. Here $ \mathcal{O}_A^{(\prime)} $ refers to the self-hermitian/complex operators respectively.

To organize the expansion of the EFT we adopt the counting of chiral dimension \cite{Buchalla:2013eza} as a generalization of the Naive Dimensional Analysis (NDA) \cite{Georgi:1992dw}  for EW effective field theories with strong underlying dynamics. It will be apparent that this counting agrees with the naive estimates for the effective operators considered in ref. \cite{Sannino:2016sfx}. In a realistic FPC model the power-counting is complicated slightly, by the potential occurrence of strong Yukawa couplings; achieving the correct top mass requires the product $ y_{Q_3} y_t \sim 4 \pi $. Strong couplings in the chiral expansion, can potentially enhance certain operators beyond the order ascribed to them by simple counting of the chiral dimension. To alleviate this issue we defined the effective Yukawa couplings 
	\begin{equation}
	\dfrac{y_\mathrm{fund}}{\sqrt{4\pi}} \to y,
	\end{equation} 
which are simple rescalings of the fundamental couplings. This will allow us to treat the Yukawa couplings as perturbative, albeit with a chiral dimension lowered to 1/2 down from 1. This is the prescription used in the remainder of this article. In the end, one must remember that the Yukawa parameters entering in the EFT, are different from the fundamental Yukawa couplings by a rescaling. We would like to emphasize that a fundamental Yukawa coupling becoming strong at the composite scale is not a problem for the fundamental model. The leading order RGE analysis indicates that the coupling can run small in the UV, thus avoiding any Landau pole issues \cite{Sannino:2016sfx}.

\subsection{Chiral Lagrangian Setup}

The effective low-energy limit of the model may be described by a non-linearly realised chiral Lagrangian, incorporating the Goldstone modes of the spontaneously broken symmetry \cite{Coleman:1969sm,Callan:1969sn}. As discussed in the previous section, the TC sector is invariant under the  $ \SU( N_\tcf ) $ flavour symmetry, which is broken to the stability group $ \Sp( N_\tcf ) $ by the fermion condensate $ \Sigma_0 $. The breaking pattern will result in $\displaystyle{ {N_\tcf (N_\tcf-1)}/{2}-1} $ broken generators $ X^{i} $ with corresponding (p)NGBs  $\Pi_i$. The associated  manifold $ \SU( N_\tcf )/ \Sp( N_\tcf ) $ is parametrised by 
	\begin{equation}
	u(x) = \exp\left[\frac{ \sqrt{2} i}{f_\TC} \Pi_i(x) X^i \right],
	\label{eq:u_def}
	\end{equation}
having normalised the generators as $ \Tr [X^i X^j] = \tfrac{1}{2} \delta^{ij} $. 
The Goldstone matrix $ u $ transforms as 
	\begin{equation}
	u \longrightarrow g u h^\dagger,
	\end{equation}
under flavour transformations, with $ g\in \SU( N_\tcf ) $. Here $ h(g,\Pi) \in \Sp( N_\tcf ) $ is  a space-time dependent element of the stability group uniquely determined via the constraint $ g u h^{\dagger} \in \SU( N_\tcf )/ \Sp( N_\tcf ) $. This results in a well defined, though highly non-trivial transformation of the NGBs. Utilising the fact that the broken generators satisfy $ X_i \Sigma_0 = \Sigma_0 X_i\transpose $, one may parametrise the low-lying, pNGB, bi-linear fermion composite states as 
	\begin{equation}
	\Sigma = u \Sigma_0 u\transpose = u^2 \Sigma_0, 
	\end{equation}
transforming like $ \Sigma \rightarrow g \Sigma g\transpose $ while leaving the vacuum alignment unchanged. This parametrisation of the pNGBs around the vacuum coincides with that of Ref. \cite{Cacciapaglia:2014uja} (even though the normalisation of the decay constant is different).

As discussed in the previous section the SM gauge symmetries are embedded into the global symmetries. Parts of these are therefore promoted to local symmetries leading to the introduction of the covariant derivative $ D_\mu$. With this gauging, the lowest order effective theory reads: 
\begin{equation} \label{eq:LTC}
	\cL_2 = \frac{1}{8} f_\TC^2\Tr \left[u_\mu u^\mu + \chi_+ \right] \ .
	\end{equation}
 Following Ref. \cite{Bijnens:2009qm} we introduced
 	\begin{align}
	u_\mu &= 4i X^{i}\; \Tr\left[X^{i} u^{\dagger} D_\mu u \right] 
	 \longrightarrow h u_\mu h^{\dagger},\\
	\chi_\pm &= u^{\dagger} \chi \Sigma_0 u^{\dagger} \pm u \Sigma_0 \chi^{\dagger} u \longrightarrow h \chi_\pm h^{\dagger},  
	\end{align}
both transforming homogeneously under the stability group. The TC-fermion mass is encoded in $\chi = 2B_0  m^{\ast}_{\tcf} $, where $ B_0 $ is a TC constant. Formally this is considered to be a spurion field which transforms as $ \chi \rightarrow g \chi g\transpose $ to preserve $ \SU( N_\tcf ) $ invariance through all steps.   
For a detailed discussion of the NLO pion Lagrangian we refer to Refs. \cite{Bijnens:2011fm, Hansen:2015yaa}. We also note that the chiral Lagrangian allows for the inclusion of a topological term, known as the Wess-Zumino-Witten term, which has been gauged in \cite{Duan:2000dy}.

\subsection{Effective Bilinear Operators with Standard Model Fermions}
We now turn to the effective operators in terms of the SM fermion fields starting with the bilinear ones. They can be neatly organised according to their  chiral dimension, starting with the lowest one which reads:
 	\begin{equation}
	\mathcal{O}_{\mathrm{Yuk}} = -\dfrac{f_\TC }{2} \ (\spur{i_1}{a_1} \spur{i_2}{a_2})\, \Sigma^{a_1 a_2} \epsilon_{i_1 i_2 }\ .
	\label{eq:OH}
	\end{equation}
The above corresponds to ordinary mass terms for the SM fermions, and contains the Higgs couplings at linear order in the pNGB fields. The anti-symmetric matrix $\epsilon_{i_1 i_2 }$, defined in Appendix~\ref{app:definitions}, contracts the $\Sp(2N_\tcs)$ indices, while spinor indices are hidden with the convention that two Weyl spinors in parenthesis are contracted to a scalar.

At the next order  we have the operator:
	\begin{equation} \label{eq:OPif}
	\mathcal{O}_{\Pi f} = \dfrac{i f_\TC}{2 \Lambda_\TC} \ (\spurbar{i_1}{a_1} \bar{\sigma}_\mu \spur{i_2}{a_2} )\  \Sigma _{a_1 a_3}^\dag  \overleftrightarrow{D}^\mu \Sigma ^{a_3 a_2}\  \epsilon _{i_1 i_2}\,,
	\end{equation}	
	The above affects the coupling of massive gauge bosons, contained in the covariant derivative, to the SM fermions. 

  At next order again we find the dipole operators:
	\begin{align}
	\mathcal{O}_{f W} &= \frac{f_\TC }{2 \Lambda_{\TC}^2  } \ (\spur{i_1}{a_1} \sigma^{\mu \nu} \spur{i_2}{a_2} ) A_{\mu\nu}^I \left( T^{I}_{\tcf} \Sigma - \Sigma (T^{I}_\tcf)\transpose \right)^{a_1 a_2} \epsilon_{i_1 i_2}, \label{eq:OfW}\\
	\mathcal{O}_{f G} &= \frac{f_\TC }{2 \Lambda_{\TC}^2 } \  (\spur{i_1}{a_1} \sigma^{\mu \nu} \spur{i_2}{a_2} ) G^{A}_{\mu\nu} \Sigma^{a_1 a_2} \left( \epsilon T_\tcs^{A} - (T_\tcs^{A})\transpose \epsilon \right)_{i_1 i_2}, \label{eq:OfG} 
	\end{align}
where $T^k_{\tcf/\tcs}$ are the generators of $\SU(N_\tcf)$ and $\Sp(2N_\tcs)$ respectively, and $ A_{\mu \nu}^k$/$G_{\mu \nu}^k $ the field strength tensors of the relative gauge bosons (more precisely, of the gauged subgroup). We note that the gauge couplings constants have been absorbed into the generators $T^k_{\tcf/\tcs}$ to account for there being several SM gauge groups embedded into each of them.  
The two operators, \eqref{eq:OfW} and \eqref{eq:OfG}, have structures mimicking the Penguin-induced operators in the SM~\footnote{The naming of these operators are loosely inspired by the corresponding operators in the SM effective field theory~\cite{Grzadkowski:2010es}.}.

\subsection{Effective Four-Fermion Operators with Standard Model Fermions} \label{sec:4fermionOps}

We now construct a consistent basis of four-fermion operators starting with five independent operators featuring two left-handed spinors $\psi$ and two right-handed ones $\bar{\psi}$:
	\begin{align}
	\mathcal{O}_{4f}^1 &= \dfrac{1}{4 \Lambda_\TC^2} (\spur{i_1}{a_1} \spur{i_2}{a_2} ) (\spurbar{i_3}{a_3} \spurbar{i_4}{a_4} ) \Sigma^{a_1 a_2} \Sigma^\dagger_{a_3 a_4} \epsilon_{i_1 i_2} \epsilon_{i_3 i_4}\ ,\label{eq:fourfermion1} \\
	\mathcal{O}_{4f}^2 &= \dfrac{1}{4 \Lambda_\TC^2} (\spur{i_1}{a_1} \spur{i_2}{a_2} ) (\spurbar{i_3}{a_3} \spurbar{i_4}{a_4} ) \left(\delta^{a_1}_{\enspace a_3} \delta^{a_2}_{\enspace a_4} - \delta^{a_1}_{\enspace a_4} \delta^{a_2}_{\enspace a_3} \right) \epsilon_{i_1 i_2} \epsilon_{i_3 i_4}\ , \\
	\mathcal{O}_{4f}^3 &= \dfrac{1}{4 \Lambda_\TC^2} (\spur{i_1}{a_1} \spur{i_2}{a_2} ) (\spurbar{i_3}{a_3} \spurbar{i_4}{a_4} ) \Sigma^{a_1 a_2} \Sigma^\dagger_{a_3 a_4} \left(\epsilon_{ i_1 i_4} \epsilon_{ i_2 i_3} - \epsilon_{ i_1 i_3} \epsilon_{ i_2 i_4} \right)\ ,	\\
	\mathcal{O}_{4f}^4 &= \dfrac{1}{4 \Lambda_\TC^2} (\spur{i_1}{a_1} \spur{i_2}{a_2} ) (\spurbar{i_3}{a_3} \spurbar{i_4}{a_4} ) \left( \delta^{a_1}_{\enspace a_3} \delta^{a_2}_{\enspace a_4} \epsilon_{ i_1 i_3} \epsilon_{ i_2 i_4} + \delta^{a_1}_{\enspace a_4} \delta^{a_2}_{\enspace a_3} \epsilon_{ i_1 i_4} \epsilon_{ i_2 i_3}\right)\ , \\
	\mathcal{O}_{4f}^5 &= \dfrac{1}{4 \Lambda_\TC^2} (\spur{i_1}{a_1} \spur{i_2}{a_2} ) (\spurbar{i_3}{a_3} \spurbar{i_4}{a_4} ) \left( \delta^{a_1}_{\enspace a_3} \delta^{a_2}_{\enspace a_4} \epsilon_{ i_1 i_4} \epsilon_{ i_2 i_3} + \delta^{a_1}_{\enspace a_4} \delta^{a_2}_{\enspace a_3} \epsilon_{ i_1 i_3} \epsilon_{ i_2 i_4}\right)\ , \label{eq:fourfermion5}
	\end{align}
where $\bar{\psi}^{a,i} = \epsilon^{i j} \bar{\psi}^a_j$. Note also that the above operators are self-conjugate.
Similarly, one can construct five corresponding operators containing four left-handed spinors. However, we find that only three of them are truly independent, as shown in Appendix~\ref{app:4f_basis}. We take these three to be:
	\begin{align}
	\mathcal{O}_{4f}^6 &= \dfrac{1}{8 \Lambda_\TC^2} (\spur{i_1}{a_1} \spur{i_2}{a_2} ) (\spur{i_3}{a_3} \spur{i_4}{a_4} )  \Sigma^{a_1 a_2} \Sigma^{a_3 a_4} \epsilon_{i_1 i_2} \epsilon_{i_3 i_4}\,, \label{eq:fourfermion6} \\
	\mathcal{O}_{4f}^7 &= \dfrac{1}{8 \Lambda_\TC^2} (\spur{i_1}{a_1} \spur{i_2}{a_2} ) (\spur{i_3}{a_3} \spur{i_4}{a_4} )  \left(\Sigma^{a_1 a_4} \Sigma^{a_2 a_3} - \Sigma^{a_1 a_3} \Sigma^{a_2 a_4}\right) \epsilon_{i_1 i_2} \epsilon_{i_3 i_4}\,,   \\
	\mathcal{O}_{4f}^8 &= \dfrac{1}{8 \Lambda_\TC^2} (\spur{i_1}{a_1} \spur{i_2}{a_2} ) (\spur{i_3}{a_3} \spur{i_4}{a_4} )  \Sigma^{a_1 a_2} \Sigma^{a_3 a_4} \left(\epsilon_{i_1 i_4} \epsilon_{i_2 i_3} - \epsilon_{i_1 i_3} \epsilon_{i_2 i_4}\right)\,.
	\end{align}	
For completeness, we also show the two-dependent operators
\begin{align}
	\mathcal{O}_{4f}^9 &= \dfrac{1}{8\Lambda_\TC^2} (\spur{i_1}{a_1} \spur{i_2}{a_2} ) (\spur{i_3}{a_3} \spur{i_4}{a_4} ) \left( \Sigma^{a_1 a_3} \Sigma^{a_2 a_4} \epsilon_{i_1 i_3} \epsilon_{i_2 i_4} + \Sigma^{a_1 a_4} \Sigma^{a_2 a_3} \epsilon_{i_1 i_4} \epsilon_{i_2 i_3}\right),\\
	\mathcal{O}_{4f}^{10} &= \dfrac{1}{8\Lambda_\TC^2} (\spur{i_1}{a_1} \spur{i_2}{a_2} ) (\spur{i_3}{a_3} \spur{i_4}{a_4} ) \left( \Sigma^{a_1 a_3} \Sigma^{a_2 a_4} \epsilon_{i_1 i_4} \epsilon_{i_2 i_3} + \Sigma^{a_1 a_4} \Sigma^{a_2 a_3} \epsilon_{i_1 i_3} \epsilon_{i_2 i_4} \right), \label{eq:fourfermion10}
\end{align}
which are related to $ \mathcal{O}^{6-8}_{4f} $ via
\begin{equation}
\mathcal{O}_{4f}^6 + \mathcal{O}_{4f}^9 = 0\,, \quad \mathcal{O}_{4f}^7 +\mathcal{O}_{4f}^8 - \mathcal{O}_{4f}^{10} = 0\,. 
\end{equation}
For the case of $N_\tcf=4$ one can write another operator:
\begin{equation}
\mathcal{O}_A  = - \dfrac{1}{8 \Lambda_\TC^2} (\spur{i_1}{a_1} \spur{i_2}{a_2} ) (\spur{i_3}{a_3} \spur{i_4}{a_4} )  \epsilon^{a_1 a_2 a_3 a_4} \epsilon_{i_1 i_2} \epsilon_{i_3 i_4}\,, ~ {\rm for~}~N_\tcf=4 \ , 
\end{equation}
where $\epsilon^{a_1 a_2 a_3 a_4}$ is the fully antisymmetric 4-index matrix which is naturally linked to the ABJ anomaly of the global ${\mathrm U}(1)_{\tcf}$. However this operator is already contained in the list above because of the following operator identity:
\begin{equation}
\mathcal{O}_A =\mathcal{O}_{4f}^{10} - \mathcal{O}_{4f}^8 - \mathcal{O}_{4f}^9   = \mathcal{O}_{4f}^6 + \mathcal{O}_{4f}^7 \ , ~ {\rm for~}~N_\tcf=4 \ . 
\end{equation} 
 
\begin{figure}
	\includegraphics{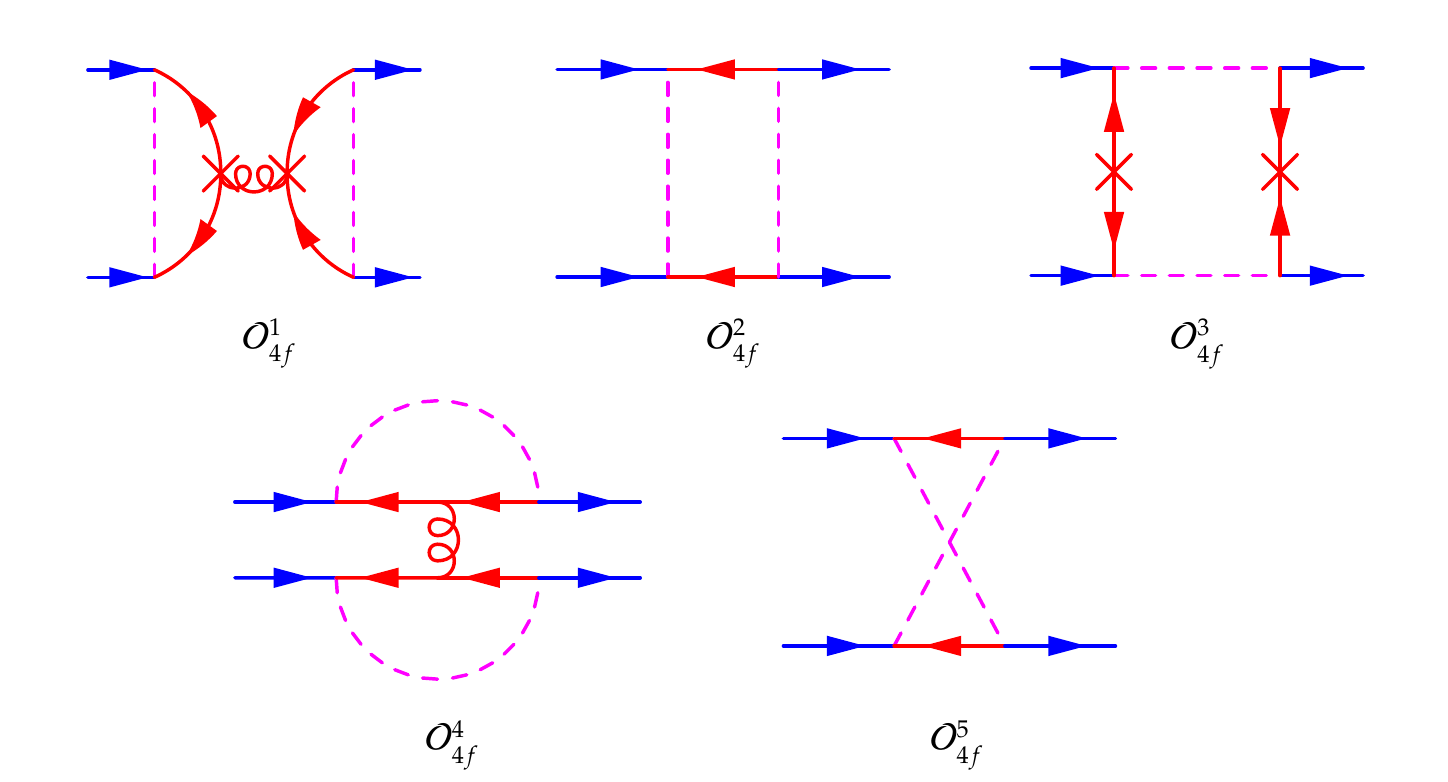}
	\caption{\emph{Representative Feynman diagrams corresponding to the operators $\mathcal{O}_{4f}^1-\mathcal{O}_{4f}^5$ in eq. (\ref{eq:fourfermion1}-\ref{eq:fourfermion5}). The blue coloured lines are SM fermions, the red coloured solid lines are TC fermions, the red coloured curly lines are TC gluons, and the magenta lines are TC scalars.}} \label{fig:diagra1}
\end{figure}

\begin{figure}
	\includegraphics{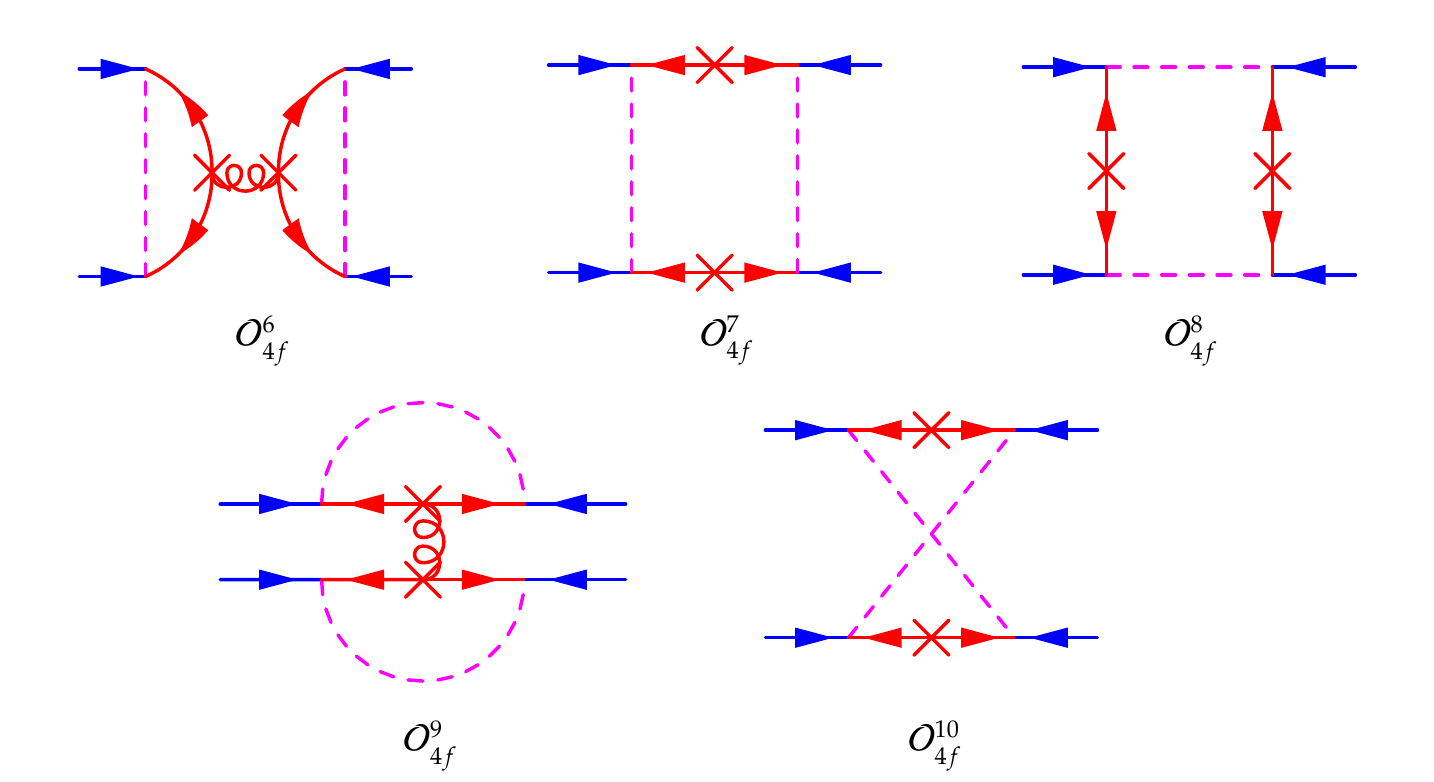}
	\caption{\emph{Representative Feynman diagrams corresponding to the operators $\mathcal{O}_{4f}^6-\mathcal{O}_{4f}^{10}$ in eq. (\ref{eq:fourfermion6}-\ref{eq:fourfermion10}). The blue coloured lines are SM fermions, the red coloured solid lines are TC fermions, the red coloured curly lines are TC gluons, and the magenta lines are TC scalars.}}\label{fig:diagra2}
\end{figure}

It is useful to represent each of the ten operators $\mathcal{O}_{4f}^{1\dots10}$ in terms of representative diagrams involving  $\tcf$ and $\tcs$ loops, as  shown in Fig.~\ref{fig:diagra1} and \ref{fig:diagra2}.  Here the ``X'' signifies an insertion of the dynamical TC-fermion mass, that is proportional to $\Sigma$. Thus the diagrams show how the $ \Sigma  $-dependence occurs in each operator. At a naive perturbative level (these diagrams are only mnemonics) the operators $\mathcal{O}_{4f}^{6-10}$  need mass insertion, while non-perturbatively one obtains operators such as $\mathcal{O}_A$  stemming from instanton corrections. 

The case in which the masses of the scalars are much heavier than $\Lambda_\TC$ is obtained by replacing
	\begin{equation}
	\epsilon_{i j} \to \Lambda_\TC^2 \left( \frac{1}{M_\tcs^2}\right)_{i j}
	\end{equation}
in each operator.  The large mass limit corresponds physically to integrating out the scalars, which in the naive diagrams corresponds to replacing each heavy scalar propagator with the inverse mass matrix. Of course one needs to identify diagrammatically the leading contributions in the inverse scalar mass expansion, as  shown in Fig.~\ref{fig:diagra1} and \ref{fig:diagra2}.

\subsection{Standard model loop-generated pNGB operators}

Loops of the elementary fermions are crucial in generating a potential for the pNGBs that includes the Higgs boson. As in other pNGB Higgs models, the potential contains radiative corrections that violate the global symmetries of the model once the spurionic Yukawa couplings assume their constant value. Accordingly, they play an important role in determining the vacuum alignment of the models.  The simplest way to write down the fermion loop generated operators is to separate the Yukawa couplings $y_f$ from the elementary fermions: the Yukawa spurions thus inherit the same quantum numbers as $\psi$ under the global symmetries of the strong dynamics, but they also acquire transformation properties under the SM gauge symmetries as carried by the elementary fermions. If a SM fermion is in the representation $R_{\SM}$ of the SM gauge group then the corresponding $y_f$  transforms as:
	\begin{equation} \label{eq:spuriony}
	(y_f)^{i}\phantom{}_{a}  \in {\tiny\yng(1)}_{\tcs}\otimes \overline{\tiny\yng(1)}_{\tcf} \otimes \overline{R}_{\SM},
	\end{equation} 
where, for simplicity, we do not explicitly write the gauge SM indices.

\subsubsection{Radiatively generated pNGB potential}
 
At leading order in the chiral expansion, and quadratic order in the spurions,  two operators might appear:
\begin{equation} \label{eq:neutrino}
\frac{f_\TC \Lambda_\TC^3}{16 \pi^2} \ (y_f)^{i_1}\phantom{}_{a_1} (y_{f'})^{i_2}\phantom{}_{a_2}\ \Sigma^{a_1 a_2} \epsilon_{i_1 i_2}\,, \quad \frac{f_\TC \Lambda_\TC^3}{16 \pi^2}\  (y_f^\ast)^{i_1,a_1} (y_{f'})^{i_2}\phantom{}_{a_2} \delta_{a_1}\phantom{}^{a_2} \epsilon_{i_1 i_2}\,.
\end{equation}
However the latter is independent on the pNGB fields and therefore just an irrelevant constant in the potential, while the former is not SM gauge invariant and therefore is not generated~\footnote{The former is due to the fact that the combination of Yukawas has the quantum numbers of mass terms for the SM fermions. Thus, the only term that may survive is proportional to the Majorana mass of right-handed neutrinos.}. 

\begin{figure}
	\includegraphics{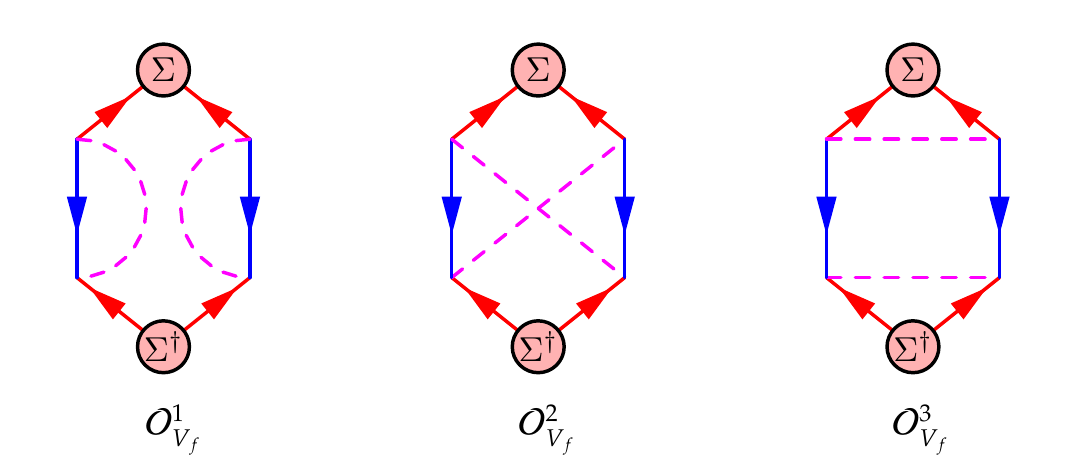}
	\caption{\emph{Representative Feynman diagrams corresponding to the operators $\mathcal{O}_{V_f}^1-\mathcal{O}_{V_f}^3$ in eq. (\ref{eq:OVf1}-\ref{eq:OVf3}). The blue coloured lines are SM fermions, the red coloured solid lines are TC fermions, and the magenta lines are TC scalars.}} \label{fig:OVf diagrams}
\end{figure}

In contrast to the lack of operators at quadratic order in the spurions, there is a plethora of operators at quartic order. They involve loops of two SM fermions, each contracting the SM indices of two spurions $ y_f $. Thus they all share the spurion structure
\begin{equation}
 (y_f^\ast y_f)^{a_1} \phantom{}_{a_2} \phantom{}^{i_1 i_2}\,, 
 \end{equation}
  where the SM indices are contracted inside the parentheses and a sum over the SM fermions $f$ is left understood. This gives rise to three operators contributing to the pNGB potential:
	\begin{align}
	\mathcal{O}^1_{V_f} = \dfrac{f_\TC^2 \Lambda_\TC^2 }{16 \pi^2} (y_f^\ast y_f)^{a_1} \phantom{}_{a_2} \phantom{}^{i_1 i_2} (y_{f'}^\ast y_{f'})^{a_3} \phantom{}_{a_4} \phantom{}^{i_3 i_4} \Sigma^{\dagger}_{a_1 a_3} \Sigma^{a_2 a_4} \epsilon_{i_1 i_2} \epsilon_{i_3 i_4}\ , \label{eq:OVf1}\\
	\mathcal{O}^2_{V_f} = \dfrac{f_\TC^2 \Lambda_\TC^2 }{16 \pi^2} (y_f^\ast y_f)^{a_1} \phantom{}_{a_2} \phantom{}^{i_1 i_2} (y_{f'}^\ast y_{f'})^{a_3} \phantom{}_{a_4} \phantom{}^{i_3 i_4} \Sigma^{\dagger}_{a_1 a_3} \Sigma^{a_2 a_4} \epsilon_{i_1 i_3} \epsilon_{i_2 i_4}\ , \label{eq:OVf2}\\
	\mathcal{O}^3_{V_f} = \dfrac{f_\TC^2 \Lambda_\TC^2 }{16 \pi^2} (y_f^\ast y_f)^{a_1} \phantom{}_{a_2} \phantom{}^{i_1 i_2} (y_{f'}^\ast y_{f'})^{a_3} \phantom{}_{a_4} \phantom{}^{i_3 i_4} \Sigma^{\dagger}_{a_1 a_3} \Sigma^{a_2 a_4} \epsilon_{i_1 i_4} \epsilon_{i_2 i_3}\ , \label{eq:OVf3}
	\end{align}     
all of which satisfy all symmetries. Again one may construct mnemonic, representative diagrams for the operators cf. Fig. \ref{fig:OVf diagrams}. The factor of $ 16\pi^2 $ is a naive effort to account for the loops of the elementary fermions. One can think of these operators as coming from the three different ways of contracting the external SM fermions in operators $ \mathcal{O}_{4f}^{1,3} $. 

As in any other composite Higgs model there are contributions to the pNGB potential stemming from SM gauge bosons. At lowest order this is due to the operator 
 	\begin{equation} \label{eq:OVg}
 	\mathcal{O}_{V_g} = \dfrac{f_\TC^2 \Lambda_\TC^2 }{16\pi^2} \Tr\left[T_\tcf^{I}\Sigma \, (T_\tcf^{I})\transpose \Sigma^{\dagger} \right]\ .
 	\end{equation}
Together with the $ \chi_+ $ term in eq. \eqref{eq:LTC} stemming from the fundamental fermion masses, the operators mentioned in this section are responsible for the pNGB potential at leading order.

\subsubsection{Radiative corrections to the kinetic terms}
At NLO in the chiral expansion one finds corrections to the pNGB kinetic terms. We find a total of 21 such operators involving the 4 $ y_f $ spurions the full list of which can be found in appendix \ref{app:pNGB_kinteic}. Physically, they give corrections to the masses of the EW gauge bosons, however, we find that only 6 of them contribute to the oblique $ T $ parameter\footnote{Assuming couplings to all SM fermions and right-handed neutrinos with fundamental Yukawa couplings as given in eq. \eqref{eq:Lfund_y}.}. They are:  
	\begin{align}
	\mathcal{O}_{y\Pi D}^{1} &= \dfrac{1}{4} \, \dfrac{f_\TC^2}{16 \pi^2} (y_f^\ast y_f)^{a_1} \phantom{}_{a_2} \phantom{}^{i_1 i_2} (y_{f'}^\ast y_{f'})^{a_3} \phantom{}_{a_4} \phantom{}^{i_3 i_4}  (\Sigma^\dagger \overleftrightarrow{D}^\mu \Sigma)_{a_1} \phantom{}^{a_2} (\Sigma^\dagger \overleftrightarrow{D}^\mu \Sigma)_{a_3} \phantom{}^{a_4} \epsilon_{i_1 i_2} \epsilon_{i_3 i_4}\ , \\
	\mathcal{O}_{y\Pi D}^{2} &= \dfrac{1}{4} \, \dfrac{f_\TC^2}{16 \pi^2} (y_f^\ast y_f)^{a_1} \phantom{}_{a_2} \phantom{}^{i_1 i_2} (y_{f'}^\ast y_{f'})^{a_3} \phantom{}_{a_4} \phantom{}^{i_3 i_4}  (\Sigma^\dagger \overleftrightarrow{D}^\mu \Sigma)_{a_1} \phantom{}^{a_2} (\Sigma^\dagger \overleftrightarrow{D}^\mu \Sigma)_{a_3} \phantom{}^{a_4} \epsilon_{i_1 i_3} \epsilon_{i_2 i_4}\ , \\
	\mathcal{O}_{y\Pi D}^{3} &= \dfrac{1}{4} \, \dfrac{f_\TC^2}{16 \pi^2} (y_f^\ast y_f)^{a_1} \phantom{}_{a_2} \phantom{}^{i_1 i_2} (y_{f'}^\ast y_{f'})^{a_3} \phantom{}_{a_4} \phantom{}^{i_3 i_4}  (\Sigma^\dagger \overleftrightarrow{D}^\mu \Sigma)_{a_1} \phantom{}^{a_2} (\Sigma^\dagger \overleftrightarrow{D}^\mu \Sigma)_{a_3} \phantom{}^{a_4} \epsilon_{i_1 i_4} \epsilon_{i_2 i_3}\ , \\
	\mathcal{O}_{y\Pi D}^{4} &= \dfrac{1}{4} \, \dfrac{f_\TC^2}{16 \pi^2} (y_f^\ast y_f)^{a_1} \phantom{}_{a_2} \phantom{}^{i_1 i_2} (y_{f'}^\ast y_{f'})^{a_3} \phantom{}_{a_4} \phantom{}^{i_3 i_4}  (\Sigma^\dagger \overleftrightarrow{D}^\mu \Sigma)_{a_1} \phantom{}^{a_4} (\Sigma^\dagger \overleftrightarrow{D}^\mu \Sigma)_{a_3} \phantom{}^{a_2} \epsilon_{i_1 i_2} \epsilon_{i_3 i_4}\ , \\
	\mathcal{O}_{y\Pi D}^{5} &= \dfrac{1}{4} \, \dfrac{f_\TC^2}{16 \pi^2} (y_f^\ast y_f)^{a_1} \phantom{}_{a_2} \phantom{}^{i_1 i_2} (y_{f'}^\ast y_{f'})^{a_3} \phantom{}_{a_4} \phantom{}^{i_3 i_4}  (\Sigma^\dagger \overleftrightarrow{D}^\mu \Sigma)_{a_1} \phantom{}^{a_4} (\Sigma^\dagger \overleftrightarrow{D}^\mu \Sigma)_{a_3} \phantom{}^{a_2} \epsilon_{i_1 i_3} \epsilon_{i_2 i_4}\ , \\
	\mathcal{O}_{y\Pi D}^{6} &= \dfrac{1}{4} \, \dfrac{f_\TC^2}{16 \pi^2} (y_f^\ast y_f)^{a_1} \phantom{}_{a_2} \phantom{}^{i_1 i_2} (y_{f'}^\ast y_{f'})^{a_3} \phantom{}_{a_4} \phantom{}^{i_3 i_4}  (\Sigma^\dagger \overleftrightarrow{D}^\mu \Sigma)_{a_1} \phantom{}^{a_4} (\Sigma^\dagger \overleftrightarrow{D}^\mu \Sigma)_{a_3} \phantom{}^{a_2} \epsilon_{i_1 i_4} \epsilon_{i_2 i_3}\ .
	\end{align}
These operators can be visualised as loops of TC-scalars and SM fermions, with TC-fermions in the external legs that close on meson fields and currents.

Again, for completeness, we note the SM gauge corrections to pNGB kinetic term. From one propagating gauge bosons, there are two operators which contribute to the $ T $ parameter: 
	\begin{align}
	\mathcal{O}_{\Pi D}^1 &= \dfrac{1}{4} \dfrac{f_\TC^2}{16 \pi^2} \Tr \left[ (\Sigma \overleftrightarrow{D}_\mu \Sigma^\dagger) T_\tcf^{I} (\Sigma \overleftrightarrow{D}^\mu \Sigma^\dagger) T_\tcf^{I}  \right] \, ,\\
	\mathcal{O}_{\Pi D}^2 &= \dfrac{1}{4} \dfrac{f_\TC^2}{16 \pi^2} \Tr \left[ (\Sigma \overleftrightarrow{D}_\mu \Sigma^\dagger) T_\tcf^{I} \right] \Tr \left[ (\Sigma \overleftrightarrow{D}^\mu \Sigma^\dagger) T_\tcf^{I}  \right]\, .
	\end{align}
Here there is an implicit sum over all the gauge bosons $ I $, and a trace over the $ \SU(N_\tcf) $ index. The full list can again be found in appendix \ref{app:pNGB_kinteic}.

Furthermore, also at NLO in the chiral expansion the operator  
	\begin{equation}
	\mathcal{O}_{WW} = \dfrac{f_\TC^2}{2\Lambda_\TC^2} A^{I}_{\mu\nu} A^{J\mu\nu} \, \Tr \left[ T^{I}_\tcf \Sigma (T^{J}_\tcf)\transpose \Sigma^\dagger \right]
	\end{equation}
gives the only contribution to the $ S $ parameter.

 \section{Top and bottom physics in the most minimal model of Fundamental Partial Compositeness}
\label{topandbottom}

We now specialise to the most minimal model~\cite{Sannino:2016sfx}, defined by  the choice of gauge group $G_\TC = \SU(2) \sim \Sp(2) $ and $N_\tcf = 4$ Weyl TC-fermions in the fundamental representation. We start the analysis by studying in detail the minimal TC-scalar sector to give mass to top and bottom alone. 
The TC-scalar sector, therefore, only contains a single field $\tcs_t$, with quantum numbers  summarised in Table~\ref{tab:TC_states-top}: the global symmetry is  $\Sp(6)$ since $N_\tcs =3$.  With respect to the SM gauge group $ G_\SM $, the Weyl TC-fermions transform as  $ \tcf_Q\in (1,2)_0 $,  $ \tcf_u \in (1,1)_{-1/2} $ and $ \tcf_d \in (1,1)_{1/2} $. The  overall theory is gauge-anomaly free. Note that the fermionic sector of this TC model was originally proposed in Refs.~\cite{Ryttov:2008xe,Galloway:2010bp}.
The vacuum alignment of the theory can be written as the following anti-symmetric matrix in the $\SU(4)$ space~\cite{Cacciapaglia:2014uja}:
	\begin{equation}
	\Sigma_0^{a b} = \left( \begin{array}{cccc}
	0 & c_\theta & s_\theta & 0 \\
	- c_\theta & 0 & 0 & s_\theta \\
	- s_\theta & 0 & 0 & - c_\theta \\
	0 & - s_\theta & c_\theta & 0 
	\end{array} \right)\,.
	\end{equation}
The angle $\theta$ parameterises the alignment of the vacuum w.r.t. the EW embedding~\cite{Dugan:1984hq}, and relates the pNGB decay constant to the EW scale as $v_{\rm EW} = f_\TC s_\theta = f_\TC \sin \theta$.

\begin{table}[t]
	\centering
	\begin{tabular}{c|ccc|ccc}
		& $\SU(3)_c$ & $\SU(2)_\LL$ & $\UU(1)_Y$ &  $ \UU(1)_B $ & $\SU(4)_\tcf$  & $\Sp (6)_\tcs$\\
		\hline
		$\f_Q$ & 1 & {\tiny\yng(1)} & 0 & & & \\
		$\f_u$ & 1 & 1 & $-\frac{1}{2} $ & 0& {\tiny\yng(1)} &  1 \\
		$\f_d$ & 1 & 1 & $\frac{1}{2}$ & &  & \\
		\hline
		$\s_t$ & $\overline{{\tiny\yng(1)}}$ & 1 & $-\frac{1}{6}$ &  $ -\tfrac{1}{3} $ & 1 &  {\tiny\yng(1)} \\
		\hline
		$Q_3$ & {\tiny\yng(1)} & {\tiny\yng(1)} & $\frac{1}{6}$ &  $\tfrac{1}{3}$ & &\\
		$u_3$ & $\overline{{\tiny\yng(1)}}$ & 1 & $-\frac{2}{3}$  &$ -\tfrac{1}{3} $ & &\\
		$d_3$ & $\overline{{\tiny\yng(1)}}$ & 1 & $\frac{1}{3}$  &$ -\tfrac{1}{3} $ & &\\
	\end{tabular}
	\caption{\emph{Fundamental technicolour states with their gauge quantum numbers and global symmetries. The table includes the 3rd generation quarks too, and the charge assignment under the baryon number $\UU(1)_B$.}}
	\label{tab:TC_states-top}
\end{table}

At the fundamental Lagrangian level the new Yukawa couplings with the SM fields read: 	
	\begin{equation}
	\mathcal{L}_{\rm {top-bottom}} = y_{Q_3} \  Q_{3,\alpha} \tcs_t \epsilon_\TC \tcf_Q^{\alpha} - y_{t} \ u_3 \tcs_t^{\ast} \tcf_d +  y_{b} \  d_3 \tcs_t^{\ast} \tcf_u  \hc \ ,
	\label{eq:Ltop_y}
	\end{equation}
where $ \alpha $ is the $ \SU(2)_\LL $ index, and $u_3$ and $d_3$ are the left-handed spinors constructed out of the charge-conjugate right-handed top and bottom singlets. The above Yukawa interactions can be written in the compact form of eq.~\eqref{eq:fund_Yukawa} by defining a spurion
	\begin{equation}
	\spur{i}{\,a}= \left( \begin{array}{cccc}
	0 & 0 & y_{b}\  d_3 & - y_{t}\ u_3 \\
	y_{Q_3}\ q_3^{(d)} & -y_{Q_3}\ q_3^{(u)} & 0 & 0
	\end{array} \right)\,,
	\end{equation}
where each row transforms as  anti-fundamental of $ \SU(4)_\tcf $ and each column as a fundamental of   $ \Sp(6)_\tcs $~\footnote{The implicit QCD colour indices of the quarks are embedded as part of $\Sp(6)$.}. Note that $ Q_{3,\alpha} = \varepsilon_{\alpha \beta} Q_3^\beta = (-q_3^{(d)}, q_3^{(u)})$ transforms as an anti-doublet of $ \SU(2)_\LL $, while $ (y_{b} d_3, -y_{t} u_3) $ as a doublet of $ \SU(2)_\RR $, consistently with the decomposition of an $ \overline{ {\tiny\yng(1)} } $ of $ \SU(4)_\tcf $.

The operator $ \mathcal{O}_{\mathrm{Yuk}} $, in eq. \eqref{eq:OH},  is responsible for the generation of the SM fermion masses and Yukawa couplings to the Higgs boson (up to effects of non-linearities in the pNGB fields):
	\begin{equation}
	\mathcal{L}_{\rm EFT} \supset 
	 -C_{\mathrm{Yuk}}\ v_\mathrm{EW}   \left( y_{Q_3} y_{b}\ q_3^{(d)} d_3 + y_{Q_3} y_{t }\ q_3^{(u)} u_3 \right)\left(1 + \dfrac{c_\theta h} {v_{\mathrm{EW}}} + \dots \right) \hc
	\label{eq:top_bot_yukawa}
	\end{equation}
The top and bottom masses can, thus, be identified with 
	\begin{equation} \label{eq:topmass}
	m_t = \abs{C_{\mathrm{Yuk}}\ y_{Q_3} y_{t}} v_\mathrm{EW} \andeq m_b = \abs{C_{\mathrm{Yuk}}\  y_{Q_3} y_{b}} v_\mathrm{EW} \ . 
	\end{equation}

A potential for the Higgs boson, and the other pNGB, generated by loops of top and bottom, is encoded in the operators in eqs.~\eqref{eq:OVf1}, \eqref{eq:OVf2} and~\eqref{eq:OVf3}. 
Expanding in the pNGB fields, the term that corresponds to a potential for the alignment angle $\theta$ reads
	\begin{equation}
	V_{\rm t/b} (\theta) = - \frac{3 f_\TC^2 \Lambda_\TC^2}{8 \pi^2} \left[ |y_{Q_3}|^2(|y_t|^2 + |y_b|^2) ( 3 C_{V_f}^1 + C_{V_f}^2)\ s^2_\theta +   (|y_{Q_3}|^4 +|y_t y_b|^2) (3 C_{V_f}^1 - C_{V_f}^3)\ c_\theta^2 \right]\,. 
	\label{Zedisdead}
	\end{equation}
This first term, proportional to $s_\theta^2$, has the same form as the contribution generated by a direct bilinear coupling of the top and bottom to the TC-fermions, as used in Ref.~\cite{Galloway:2010bp,Cacciapaglia:2014uja}: the combinations of Yukawas, in fact, are proportional to the top and bottom masses.
As usual, expecting a negative sign in front coming from the fermion loop, this term alone tends to destabilise the vacuum alignment towards the TC limit $\theta = \pi/2$.
The second term, proportional to $c_\theta^2$, is new in FPC models and, depending on the sign of the coefficients, it may either contribute to the destabilisation or tend to flip the alignment to the EW preserving direction. To achieve electroweak symmetry breaking one should have $|y_{Q_3}| < |y_t|$, which, as we shall see, is supported by the constraints coming from the $ Z $ boson to $\bar{b} b$. 

The potential also receives contributions from the gauge interactions, encoded in eq.~\eqref{eq:OVg}, and the TC-fermion mass, as shown in eq.~\eqref{eq:LTC}, which have the same form as in models without FPC~\cite{Galloway:2010bp,Cacciapaglia:2014uja,Katz:2005au}. In particular, the contribution of the TC-fermion mass can be used to stabilise the potential around small $\theta$ values against the top loops, in order to obtain a pNGB Higgs.
Note that higher dimension operators generated by top loops may also help stabilising the potential, however they are expected to be subleading.

\subsection{Couplings of the $Z$ to the bottom quark} \label{sec:Zbb}

We now turn to the operator in eq.~\eqref{eq:OPif}, that generates corrections to the gauge couplings of the massive gauge bosons to fermions:
\begin{multline} \label{eq:Zbb}
\mathcal{O}_{\Pi f} =  \frac{g}{2 \cos \theta_W} \dfrac{f_\TC}{\Lambda_\TC} s^2_\theta\;  Z_\mu \left( |y_{Q_3}|^2\ (\bar{t}_\LL \gamma^\mu t_\LL - \bar{b}_\LL \gamma^\mu b_\LL) + |y_{b}|^2\  \bar{b}_\RR \gamma^\mu b_\RR - |y_{t}|^2\ \bar{t}_\RR \gamma^\mu t_\RR) \right) \\
- \frac{g}{\sqrt{2}} \dfrac{f_\TC}{\Lambda_\TC} s^2_\theta\;  W^+_\mu \left( y_{b}^\ast y_{t}\ \bar{t}_\RR \gamma^\mu b_\RR - |y_{Q_3}|^2\ \bar{t}_\LL \gamma^\mu b_\LL \right) \hc
\end{multline}
where the SM top and bottom are in the usual Dirac spinor notation.
While the couplings of the top to the $ Z $ are unconstrained, and $y_b$ can be taken small to reproduce the bottom mass, the coupling of the left-handed bottom to the Zed receives sizeable corrections proportional to $|y_{Q_3}|^2$. The well known issue is that $y_{Q_3}$ coupling cannot be too small, as it enters the formula for the top mass.
Imposing the latest constraints~\cite{Baak:2014ora,Gori:2015nqa}, we obtain the $2\sigma$ limit~\footnote{For all our numerical estimates we have used $ \Lambda_\TC = 4\pi f_\TC $.}~\footnote{ Please note that all bounds found here, are on the effective rather than the fundamental Yukawa parameters.}: 
\begin{equation}
 C_{\Pi f} |y_{Q_3}|^2 s^2_\theta < 0.043\,, \qquad \mbox{@ 95\% CL}\,. 
\end{equation}
This constraint mainly comes from the measurement of $R_b$ at LEP~\cite{ALEPH:2005ab}.
The constraint on $\theta$ from electroweak precision tests tends to ease the tension, as $ s^2_\theta \lesssim 0.1$ is generically required~\cite{Arbey:2015exa}. Furthermore, it is possible to obtain the correct top mass with a small $y_{Q_3}$ by maximising the right-handed mixing $y_t$, i.e. assuming that the right-handed top is more composite than the left-handed part.
Interestingly, this configuration is also preferred in the top-loop induced potential for the alignment of the vacuum, as we have seen in \eqref{Zedisdead}. 
Using eq.~\eqref{eq:topmass}, the above bound translates into the following lower bound on the right-handed top mixing~\footnote{Note that our normalisation for the pre-Yukawa couplings differs from the one usually considered in EFT realisations, see Section~\ref{sec:EFT} for more details.}:
\begin{equation} \label{eq:boundZbb}
|y_t| \frac{\abs{C_{\rm Yuk}}}{\sqrt{C_{\Pi f}}} \gtrsim \frac{m_t}{f_\TC} \frac{1}{\sqrt{0.043}} = \frac{10~\mbox{TeV}}{\Lambda_\TC}\,,
\end{equation}
which is, therefore, a mild constraint at reasonable condensation scales ($\Lambda_\TC = 10$ TeV roughly corresponds to $f_\TC \simeq 3 v_\mathrm{EW}$).

\subsection{Effective interactions for the top sector}
The effective Lagrangian for EW physics contains four fermion interactions which are induced by the underlying strong dynamics. In Section~\ref{sec:4fermionOps}, we showed that there are 8 independent operators, 5 of which are self-hermitian.  
Expanding the operators $ \mathcal{O}_{4f}^i $ we obtain  four-fermion interactions involving the SM fermions listed in Appendix~\ref{app:list}. Note that these set of operators  cannot be directly matched to the Warsaw basis~\cite{Grzadkowski:2010es} because our theory contains non-linearities in the Higgs field. Effectively, this gives us the Wilson coefficient for each operator in terms of the fundamental Yukawa couplings, the scale of strong dynamics $ \Lambda_\TC $, and the coefficients $ C_{4f}^i $ of the strong dynamics. 

The phenomenologically relevant operators involve four tops, as they are directly probed at the LHC in four top final states, such as 
	\begin{equation}
	\mathcal{L}_{\mathrm{EFT}} \supset \dfrac{C_{4f}^4 + C_{4f}^5}{4 \Lambda^2_\TC } \abs{y_{t}}^4 (\bar{t}_\RR \gamma^{\mu} t_\RR) (\bar{t}_\RR \gamma_{\mu} t_\RR)  = \dfrac{C_{4f}^4 + C_{4f}^5}{4 \Lambda^2_\TC } \abs{y_{t}}^4 O_{uu}^{3333}, 
	\end{equation}
where the four $3$'s refer to the generation of each of the four fermions. ATLAS~\cite{ATLAS:2016btu} puts an upper limit on this operator at 95\% CL, yielding the constraint:
	\begin{equation}
	\dfrac{\abs{C_{4f}^4 + C_{4f}^5}}{4 \Lambda^2_\TC } \abs{y_{t}}^4 < 2.9~\mathrm{TeV}^{-2} \quad \Rightarrow \quad \abs{C_{4f}^4 + C_{4f}^5}^{1/4} |y_t| < 5.8\, \left( \frac{\Lambda_\TC}{10~\mbox{TeV}} \right)^{1/2}\,, \quad \mbox{@ 95\% CL}\,.
	\end{equation}
The above upper bound is compatible with the lower bound in eq.~\eqref{eq:boundZbb}, and the situation improves significantly for increasing values of $\Lambda_\TC$.

In addition to the four fermion interactions, the operators $ \mathcal{O}_{fW}$ and $ \mathcal{O}_{fG} $, in eqs. \eqref{eq:OfW} and \eqref{eq:OfG}, give rise to new dipole interactions between gauge fields and SM fermions. 
Knowing that the SM gauge bosons are embedded in the two global symmetries  $\SU(4)_\tcf$ and $\Sp(6)_\tcs$ in the following way:
	\begin{equation}
	A_\mu^{I} (T^{I}_{\tcf})^{a}_{\enspace b} = \frac{1}{2} \begin{pmatrix} g\ W_\mu^{i} \tau^{i} & 0 \\ 0 & -g'\ B_\mu \tau^{3} \end{pmatrix} \andeq G^{A}_\mu (T^{A}_{\tcs})^{i}_{\enspace j } = \dfrac{g_S}{2} G^{A}_\mu \begin{pmatrix} - \lambda^{\ast A} & 0 \\ 0 & \lambda^{A}\end{pmatrix} + \dfrac{g'}{6} B_\mu \begin{pmatrix} - \mathds{1} & 0 \\ 0 & \mathds{1} \end{pmatrix}\,,
	\end{equation}
where $W$, $B$ and $G$ represent respectively the $\SU(2)_\LL$, hypercharge and QCD gauge bosons respectively, 
the operators generate the following couplings:
	\begin{align}
	\mathcal{O}_{fW} 
	& = \dfrac{-1}{C_\mathrm{Yuk} \Lambda_\TC^2} \dfrac{m_t}{2\sqrt{2} v_\mathrm{EW} } \left(g \mathcal{O}^{33\ast}_{uW} + g' \mathcal{O}^{33\ast}_{uB} \right) + \dots \label{eq:OfW_simp} \\ 
	\mathcal{O}_{fG} 
	&= \dfrac{-1}{C_\mathrm{Yuk} \Lambda_\TC^2} \dfrac{\sqrt{2} m_t}{ v_\mathrm{EW} } \left( g_s \mathcal{O}^{33\ast}_{uG} + \frac{g'}{6} \mathcal{O}^{33\ast}_{uB} \right) + \dots
	\end{align}
where the Yukawa couplings have been expressed in terms of the physical top mass, as in eq.~\eqref{eq:topmass}.
The dots contain couplings of the pNGBs generated by the non-linearities, and the operators $\mathcal{O}^{33}_{uV}$ are from the SM EFT \cite{Durieux:2014xla}.
The {\tt TopFitter} Collaboration \cite{Buckley:2015lku} has extracted constraints on the anomalous couplings of the top quarks, in the EFT language, by considering the latest data on top production cross sections and distributions. The bound on $ \mathcal{O}^{33}_{uB} $ is weaker than that on $ \mathcal{O}^{33}_{uW} $,  so we can use the latter to impose bounds on $C_{fW}$: 
	\begin{equation}
	 \abs{\dfrac{C_{fW}}{C_\mathrm{Yuk}} } <   2500\ \left( \dfrac{\Lambda_\TC}{10~\mbox{TeV}} \right)^2\, \qquad \mbox{@ 95\% CL}\,.  
	\end{equation}  
The bound from the gluon coupling $ \mathcal{O}^{33}_{uG} $ yields a stronger bound~\footnote{The bounds come from the 95\% CL limits on the SM EFT operator coefficients $\dfrac{v^2_{EW}}{\Lambda^2}\abs{C_{uW}^{33}} < 0.242$ and $\dfrac{v^2_{EW}}{\Lambda^2}\abs{C_{uG}^{33}} < 0.079$ from Ref.~\cite{Buckley:2015lku}.}:
	\begin{equation}
	\abs{\dfrac{C_{fG}}{C_\mathrm{Yuk}} } <  110\ \left( \dfrac{\Lambda_\TC}{10~\mbox{TeV}} \right)^2\, \qquad \mbox{@ 95\% CL}\,.    
	\end{equation}
Both of these constraints are obtained from marginalised bounds on the operators. Limiting other operators may therefore lead to stronger bounds.

\subsection{Extension to light generations and leptons} \label{sec:light_gens}

The fundamental Lagrangian can be expanded to include all three generations of quarks and leptons. The minimal strategy~\cite{Sannino:2016sfx} is to extend the TC-scalar sector by three extra un-coloured scalars $\tcs_l$ to couple to the three generations of leptons, and two extra coloured scalars $\tcs_u$ and $\tcs_c$ (corresponding to 6 complex scalars) to couple to the two light quark generations. In total, therefore, we have $N_\tcs = 12$ complex scalars, which enjoy a global $\Sp(24)_\tcs$ symmetry. The quantum numbers of both the TC and the SM fields are summarised in Table~\ref{tab:TC_states}. 

\begin{table}
	\centering
	\begin{tabular}{c|ccc|cccc}
		& $ \SU(3)_c $ & $ \SU(2)_\LL $ & $ \UU(1)_Y $ & $ \UU(1)_B $ & $ \UU(1)_\ell $ & $ \UU(3)_{g_1} $ & $ \UU(3)_{g_2} $\\
		\hline
		$\f_Q$ & 1 & {\tiny\yng(1)} & 0 & & & & \\
		$\f_u$ & 1 & 1 & $-\frac{1}{2} $ & 0 & 0 &  1 & 1\\
		$\f_d$ & 1 & 1 & $\frac{1}{2}$ & &  & & \\
		\hline
		$\s_q$ & $\overline{{\tiny\yng(1)}}$ & 1 & $-\frac{1}{6}$  & $ -\tfrac{1}{3} $ & 0 & {\tiny\yng(1)} & 1 \\
		$\s_l$ & 1 & 1 & $\frac{1}{2}$ &0 & $ -1 $ & 1 & {\tiny\yng(1)}  \\
		\hline
		$Q$ & {\tiny\yng(1)} & {\tiny\yng(1)} & $\frac{1}{6}$ & $ \tfrac{1}{3} $ & 0 &&\\
		$u$ & $\overline{{\tiny\yng(1)}}$ & 1 & $-\frac{2}{3}$ & $ -\tfrac{1}{3} $ & 0 &&\\
		$d$ & $\overline{{\tiny\yng(1)}}$ & 1 & $\frac{1}{3}$ & $ -\tfrac{1}{3} $ & 0 &&\\
		$L$ & 1 & {\tiny\yng(1)} & $-\frac{1}{2}$ & 0 & $ 1 $&& \\
		$e$ & 1 & 1 & $-1$ & 0 & $ -1 $ &&\\
		$\nu$ & 1 & 1 & $0$ & 0 & $ -1 $ && 
	\end{tabular}
	\caption{\emph{Fundamental technicolour states and SM fermions with their SM gauge quantum numbers. The table also includes the charge assignments under the baryon and lepton number $ \UU(1)_{B,\ell} $.}} 
	\label{tab:TC_states}
\end{table}

The complete Yukawa interactions now read~\footnote{Note that the scalars are in the conjugate representation of $ G_\SM $ as compared to the minimal model suggested in Ref.~\cite{Sannino:2016sfx}.}:
\begin{multline}
	\mathcal{L}_{\mathrm{yuk}} =  y_Q\  Q_{\alpha} \tcs_q\epsilon_{\TC} \tcf_Q^{\alpha} -  y_u\  u \tcs_q^{\ast} \tcf_d +  y_d\  d \tcs_q^{\ast} \tcf_u + \\
	 y_\ell\  L_\alpha \tcs_l \epsilon_{\TC} \tcf_Q^\alpha -  y_\nu\  \nu \tcs_l^{\ast} \tcf_d +  y_e\  e \tcs_l^{\ast} \tcf_u -\tilde{y}_\nu\  \nu \tcs_l \epsilon_\TC \tcf_u \hc
	\label{eq:Lfund_y}
	\end{multline}
where each coupling is a $3\times 3$ matrix in flavour space, and the flavour indices are left implicit for readability. Table~\ref{tab:TC_states} also contains the symmetries $ U(3)_{g_{1,2}} $ corresponding to global approximate flavour symmetries between the 3 generations of each TC scalar. Additionally the full model still preserves a Baryon number symmetry as does the SM. However, the lepton number symmetry is explicitly violated by the coupling $ \tilde{y}_\nu $, not surprisingly as the inclusion of such a coupling gives rise to a Majorana mass term for the right-handed neutrinos.   

Just as in the case of the top and bottom, the Yukawa interactions can be written in the more compact form from eq. \eqref{eq:fund_Yukawa} by defining the spurion field $\psi$ as (colour and generation indices are, once again, left implicit)
\begin{equation}
	\spur{i}{a}=\begin{pmatrix}
	0 & 0 & y_d d & -y_u u \\
	0 & 0 & y_e e & -y_\nu \nu \\
	y_Q q^{(d)} & - y_Q q^{(u)} & 0 & 0 \\
	y_\ell l^{(e)} & - y_\ell l^{(\nu)} & \tilde{y}_\nu \nu & 0
	\end{pmatrix},
	\label{eq:spur_light_gens}
\end{equation}
where $a\in \SU(4)_\tcf$ and  $i\in \Sp (24)_\tcs$. Details of this construction are found in Appendix \ref{app:definitions}. The hierarchy of the fermion masses can be encoded either in the fundamental Yukawa couplings or in a hierarchy in the mass spectrum of the TC-scalars.
The phenomenology of the two scenarios is different for the low energy flavour observables as well as for the spectrum of the massive composite states of the theory.
It's noteworthy that, thanks to the compact spurion form, the effect of the light generations can be expressed in terms of the same operator basis we used for the top/bottom case. Of course, at the EW scale the effect of light quarks will be negligible, as they are suppressed by the small effective Yukawas (or scalar masses), and we leave the effects on low energy flavour physics and lepton masses for further investigations. 

\begin{figure}[t]
	\includegraphics{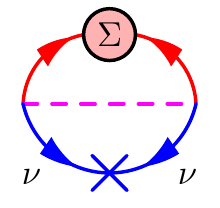}
	\caption{\emph{Representative diagram for the contribution to the Higgs potential of a right-handed neutrino with an elementary Majorana mass, symbolized here by a blue cross.}}
	\label{fig:Vnu}
\end{figure}

The only exception is given by the physics of the right-handed neutrinos that might have Majorana masses and order-1 fundamental Yukawa couplings. Note that the presence of both Yukawas $y_\nu$ and $\tilde{y}_\nu$ will also generate a composite Majorana mass for the right-handed neutrino of the order  $C_{\rm Yuk} c_\theta f_\TC y_\nu \tilde{y}_\nu$. At the same time, the first operator in eq.~\eqref{eq:neutrino} gives rise to a non-vanishing contribution to the Higgs potential
\begin{equation}
V_\nu (\theta) \sim - \dfrac{ f_\TC \Lambda_\TC^3 }{8 \pi^2} y_\nu \tilde{y}_\nu c_\theta\,,
\end{equation}
which only exists if an elementary Majorana mass is present. A mnemonic diagram for this operator is sketched in Fig. \ref{fig:Vnu}. This term has the same dependence on the alignment angle $\theta$ as the contribution of the TC fermion mass~\cite{Cacciapaglia:2014uja}, thus it can be used to stabilise the potential generated by the top loops towards small values of $\theta$ if the Yukawa couplings of at least one neutrino are of order 1. This would provide a new mechanism where partial compositeness for neutrinos generates both TeV-scale see-saw and stabilises the Higgs potential.

\section{Connections with other approaches to partial compositeness}\label{Connection}

In this section we sketch the connection between our analysis, and other approaches used in the literature to study partial compositeness. We first address effective approaches, based either on the construction of an EFT or on extra dimensional implementations. Finally, we comment on the possible applicability of our results to purely fermionic underlying theories featuring partial compositeness.

\subsection{Effective Operator approach} \label{sec:EFT}

The most popular approach to composite Higgs models in the literature has been to construct EFTs simply based on the symmetry breaking patterns (see Refs.~\cite{Marzocca:2012zn,Panico:2015jxa} for a pedagogical introduction), without any reference to the underlying theory~\footnote{This approach might be the only available one if the underlying theory is conformal, in which case it can only be defined in terms of operators and their conformal dimensions.}. 
As a consequence, to implement partial compositeness, the choice of the representation under which the top partners transform has been arbitrary. 
Furthermore, top partners in the EFT approach have been assumed to be the main driving force in the stabilisation of the vacuum alignment along the small-$\theta$ limit: this mechanism can only work if the top partners are light~\cite{Redi:2012ha,Pomarol:2012qf} and the contribution to the pNGB potential is dominated by their loops.  Accepting the lightness of top partners with respect to the natural resonance scale, i.e. $\Lambda_\TC \sim 4 \pi f_\TC$, one is justified to include them in the EFT construction. Note however that top partners are not necessarily the only contributors to the Higgs potential~\cite{Katz:2005au,Galloway:2010bp,Cacciapaglia:2014uja}.

In the case under study in this work, the representation of the top partners is fixed to be the fundamental of the global symmetry $\SU (4)$. This choice has been considered problematic in the literature, as it typically leads to large corrections to the $Z$ coupling to bottoms. However, as we will see shortly, this problem only applies if the top partners are light. It is instructive to compare our general operator approach presented in Section~\ref{topandbottom} with the results one would obtain by adding the top partners to the EFT. The couplings of the top partners, that we collectively call $\mathcal{B}$, to the SM fermions can be written as:
\begin{equation}
\mathcal{L}_{\rm PC} = - \bar{y}_{Q_3}^{\rm EFT} f_\TC \ \bar{\psi}_{Q_3} \cdot \Sigma^\dagger \cdot \mathcal{B}_R - \bar{y}_t^{\rm EFT} f_\TC\ \bar{\mathcal{B}}_L \cdot \Sigma \cdot \psi_t\,, \quad \mbox{with}\;\; \psi_{Q_3} = \left( \begin{array}{c} Q_3 \\ 0 \\ 0 \end{array} \right)\;\; \mbox{and} \;\; \psi_t = \left( \begin{array}{c} 0 \\ u_3 \\ 0 \end{array} \right)\,,
\end{equation}
where the SM fermions are embedded into spurions transforming as the fundamental of $ \SU(4)_\tcf $. The symmetries associated to the scalars $\tcs$ are thus ignored. The mass of the top can be obtained by diagonalising the resulting mass matrix, yielding
\begin{equation}
m_t = 2 M_\mathcal{B} s_\theta\ \frac{\bar{y}^{\rm EFT}_{Q_3} f_\TC}{\sqrt{M_\mathcal{B}^2 + {\bar{y}^{\rm EFT}_{Q_3}}^2 f_\TC^2}} \frac{\bar{y}^{\rm EFT}_t f_\TC}{\sqrt{M_\mathcal{B}^2 + {\bar{y}^{\rm EFT}_t}^2 f_\TC^2}} + \dots
\end{equation}
where the dots stand for higher orders in an expansion for small $s_\theta$. This equation should be compared to eq.~\eqref{eq:topmass}. We see that the two results coincide once we identify
\begin{equation} \label{eq:EFTmatching}
y_{Q_3/t} \dfrac{\sqrt{f_\TC} }{\sqrt{\Lambda_\TC}} \to \frac{\bar{y}^{\rm EFT}_{Q_3/t} f_\TC}{\sqrt{M_\mathcal{B}^2 + {\bar{y}^{\rm EFT}_{Q_3/t}}^2 f_\TC^2}}\,, \quad C_{\rm Yuk} \Lambda_\TC \to 2 M_\mathcal{B}\,. 
\end{equation}
We see that the operator estimate matches if the mass of the top-partners is at its natural value $M_\mathcal{B} \sim \Lambda_\TC$.
The mixing between SM fermions and top partners induces corrections to the gauge couplings of the top and bottom to the massive $W$ and $Z$ too, due to the fact that the top partners are vector-like fermions~\cite{delAguila:2000rc}. In the bottom sector, we thus obtain:
\begin{equation}
\frac{g}{2 \cos \theta_W}  s^2_\theta\, Z_\mu  \frac{{\bar{y}^{\rm EFT}_{Q_3}}^2 f^2_\TC}{M_\mathcal{B}^2 + {\bar{y}^{\rm EFT}_{Q_3}}^2 f_\TC^2} \bar{b}_L \gamma^\mu b_L + \dots
\end{equation}
which nicely compares with eq.~\eqref{eq:Zbb} once the identification in eq.~\eqref{eq:EFTmatching} is taken into account.
We see, therefore, that the approach with top partners in the EFT gives the same results as the effective operators we consider, and the two actually coincide if the mass of the top partners is at the natural scale $\Lambda_\TC$. Thus, for heavy top partners, the bound from the $Z$ coupling are not problematic, as we showed in Section~\ref{sec:Zbb}.

Another effective approach to partial compositeness relies on extra dimensions: it is mainly based on adapting the conjectured correspondence of anti-de-Sitter (AdS) space-time with 4-dimensional conformal field theories~\cite{Maldacena:1997re} to non-supersymmetric scenarios. Models based on warped extra dimensions have been used to characterise composite Higgses based on a conformal underlying theory~\cite{Contino:2003ve,Agashe:2004rs}. The light Higgs is  identified with an additional polarisation of gauge fields in the bulk, thus borrowing many similarities from Gauge-Higgs unification models~\cite{Hosotani:1983xw,Antoniadis:2001cv} (see also Ref.~\cite{Hosotani:2005nz} in warped space).  
The mechanism of partial compositeness is described by fermions propagating in the bulk of the extra dimensions, as discussed in Refs.~\cite{Contino:2004vy,Cacciapaglia:2008bi}.
An extra-dimensional version of the model under study can be easily obtained by promoting the global symmetries $\SU(N_\tcf) \times \Sp (2 N_\tcs)$ to gauge symmetries in the bulk, broken by boundary conditions to the SM on the Planck brane, while on the TeV brane the breaking induced by the fermion condensate, i.e. $\SU (N_\tcf) \to \Sp (N_\tcf)$, is imposed. Composite fermions are represented by bulk fermions transforming as the bi-fundamental of the symmetries, while the mixing of the SM fermions, at the basis of partial compositeness, comes from explicit mass mixings on the Planck brane~\cite{Scrucca:2003ra}.
The theory would thus automatically describe spin-1 resonances in the form of Kaluza-Klein resonances of the gauge bosons.
The advantage of extra dimensions, which is also their limitation, is the fact that the spectrum is determined by the geometry. In the model under consideration, which is not conformal in the UV, the spectrum will hardly match the prediction of a warped extra dimension.

\subsection{Pure fermionic extensions}
Traditional approaches hope to achieve partial compositeness via pure underlying gauge-fermion realisations. In this case the new composite fermion operators ${\cal B}$, that couple linearly to the SM fermions, must be built out of the underlying gauge-fermion dynamics. This necessarily limits its underlying composition. In addition the need to have the composite fermion operator ${\cal B}$ with a physical dimension such that the operator $ \Psi  {\cal B}$ (with $\Psi$ a generic SM fermion) is either super-renormalisable or marginal further constrains the underlying origin of ${\cal B}$.     Therefore one can schematically build ${\cal B}$  as follows:
\begin{equation}
	{\cal B} \sim {\cal F} {\cal F} {\cal F}, \quad {\cal F} {\cal F} {\cal X}, \quad {\cal F} {\cal X} {\cal X}, \quad {\cal F} {\cal X} {\cal Z}, \quad {\cal F} \sigma^{\mu \nu}{\cal G}_{\mu \nu}, 
\end{equation}
with $ {\cal X} $ and ${\cal Z}$ potentially new TC-fermions transforming according to different representations of the gauge group and ${\cal G}_{\mu \nu} $ the technicolour field strength. Clearly which technicolour invariant composite operator can actually be built depends on the underlying dynamics. Theories in which ${\cal B}$ is made by an even larger number of fermionic degrees of freedom are strongly disfavoured because of the anomalously large anomalous dimensions that the composite fermion must have for $\Psi {\cal B}$ to be at least a marginal operator. In fact, in \cite{Pica:2016rmv} it has been argued that even realisations with three underlying fermions are challenging~\footnote{The remaining challenge is to build a theory that actually generates the operator $\Psi {\cal B}$ with the required hierarchies for the SM fermions.}. 

 As noted in \cite{Sannino:2016sfx}  because any purely fermionic extension~\cite{Barnard:2013zea,Ferretti:2013kya, Vecchi:2015fma, Ferretti:2014qta} is required to have composite baryons with dimensions close to $5/2$, these baryons would presumably behave as if they were made by a fermion and a composite scalar similar to ours (see also~\cite{Caracciolo:2012je} for a supersymmetric realisation).  Naively, at some intermediate energy, our description can be viewed as an effective construction of the purely gauge-fermionic one with 
 
 \begin{equation}
{\cal F}(\Phi) \sim \	{\cal B} \sim {\cal F}( { \cal F} {\cal F}), \quad {\cal F} ({\cal X} {\cal F}), \quad {\cal F} ({\cal X} {\cal X}), \quad {\cal F} ({\cal X} {\cal Z}). 
\end{equation}
 Obviously this identification is just a mnemonic and it means that the composite baryon made by $\tcf \Phi$ can describe, at an intermediate effective level, one of the composite baryons with the same quantum number and physical dimensions. A similar relation can be thought for the ${\cal F} \sigma^{\mu \nu}{\cal G}_{\mu \nu}$ operator.  
  
We can use group theory to investigate related theories.  For example, from Table I of \cite{Belyaev:2016ftv},  we learn that model $M_6$, that features five two-index antisymmetric $\tcf$  under the technicolour gauge group $\SU(4)$ as well as three Dirac fermions in the fundamental representation ${\cal X}$~\cite{Ferretti:2014qta},  gives rise to composite baryons $\tcf {\cal X} {\cal X}$ and   $\tcf \overline{{\cal X}}  \overline{{\cal X}}$. At intermediate energies these composite baryons can be mapped into a fundamental partial composite theory featuring the same $\tcf$ fermions and six two index antisymmetric TC-scalars.

\section{Conclusion}\label{Concluisons}

We built consistent extensions of the standard model of fundamental partial composite nature and  determined their electroweak effective theories in terms of the standard model fields. 
The bases of effective operators of different mass dimensions were built and constrained using the symmetries of the underlying  theories.  
Our results can now be used as a stepping stone to undertake studies both in the lepton and quark flavour observables within a controlled theory of composite dynamics.

To elucidate the power of our approach, we focused on the most minimal theory of fundamental partial compositeness.
We analysed the physical consequences for the composite Higgs sector as well as the third generation quarks. 
Here we discovered new contributions to the Higgs potential generated from the left-handed mixing of top and bottom. Intriguingly, we also discovered that right-handed neutrinos with TeV scale composite Majorana masses can  affect the Higgs potential with relevant consequences for the vacuum alignment of the theory.
We  show that constraints on the top and bottom sectors can be naturally abided.
Our effective operators are ready to be deployed for full scale analyses of composite lepton and light quark flavour physics.

Finally, we provided relations with other approaches. The overall methodology can be employed to derive effective operators  stemming from related underlying composite theories of dynamical electroweak symmetry breaking able to give masses to the standard model fermions.

\section*{Acknowledgements}

We thank C.Englert and M.Russell, and the {\tt TopFitter} collaboration, for help in providing the numerical results of their latest fits, and H.Cai for contributing in the early stages of the project. 
We also acknowledge the support of the ``Institut Franco-Danois'' that allowed us to initiate this project.
G.C. acknowledges partial support from the Labex- LIO (Lyon Institute of Origins) under grant ANR-10-LABX-66 and FRAMA (FR3127, F\'ed\'eration de Recherche ``Andr\'e Marie Amp\`ere''), and thanks the CP3-Origins Institute for hospitality during the completion of this work.
H.G., F.S. and A.E.T. acknowledges partial support
from the Danish National Research Foundation grant DNRF:90.

\appendix

\section{Definitions and notation} \label{app:definitions}

Whenever we write an invariant of an $ \Sp(M) $ group, be it $\epsilon_\TC$ for $ \Sp(2N) $, $ \epsilon $ for $ \Sp(2N_\tcs) $, or $ \varepsilon $ for $ \SU(2)_\LL $, they are defined in a similar manner. 
For all three $ \epsilon $'s we define 
\begin{equation}
\epsilon_{ij} = - \epsilon^{ij} = \begin{pmatrix} 0 & -\mathds{1} \\ \mathds{1} & 0	\end{pmatrix},
\end{equation}
where $ \mathds{1} $ is a unitary matrix or $ 1 $ depending on the group. According to usual convention we take all 'up'-indices to be in the fundamental representation of a given group and 'down'-indices are taken to be in the anti fundamental. For the pseudoreal groups the epsilons can be used to raise or lower indices accordingly. Take e.g. the scalar field from equation \eqref{eq:Phi} transforming in the fundamental of $ \Sp(2N) $
\begin{eqnarray}
\Phi^{c\, i}=
\begin{pmatrix}
\tcs^c\\ -\epsilon_\TC^{cd} \tcs_d^*
\end{pmatrix}.
\end{eqnarray}
We note that when using the conjugate spurion field, we always use it transforming in the fundamental of $ \Sp(2N_\tcs) $, viz.
\begin{equation}
\spurbar{i}{a} = \epsilon^{ij}\overline{\psi}_{j}\phantom{}^{a} = \epsilon^{ij}(\spur{j}{a})^\ast.
\end{equation}

To construct the spurion field $ \psi $ of the SM fermions and Yukawa couplings from the fundamental Yukawa terms, one simply embeds the TC-scalars and TC-fermions in $ \tcf $ and $ \Phi $ respectively. Then it is simply a matter of matching the Yukawa terms to the explicitly symmetric construction in eq. \eqref{eq:fund_Yukawa}. In the case of the full model presented in section \ref{sec:light_gens} we have 
\begin{eqnarray}
\tcf^{a} = \begin{pmatrix}
\tcf_{Q_u}\\ \tcf_{Q_d}\\ \tcf_u \\ \tcf_d
\end{pmatrix},
\andeq
\Phi^{i} = \begin{pmatrix}
\tcs_q\\ \tcs_l\\ -\epsilon_\TC \tcs_q^*\\ -\epsilon_\TC \tcs_l^*
\end{pmatrix}.
\end{eqnarray}
in which case one recovers the spurion field given in eq. \eqref{eq:spur_light_gens}. 

For the definition of the $ \sigma $ matrices (and general Weyl-spinor algebra) we follow the notation in \cite{Dreiner:2008tw} where 
$\sigma^\mu$ and $\bar\sigma^\mu$ are defined as
\begin{equation}
\begin{split}
\sigma^\mu &=(1 ,\vec\sigma),\qquad \bar\sigma^\mu =(1, -\vec\sigma)
\end{split}
\end{equation}
and $\sigma^{\mu\nu} $ is defined as
\begin{equation}
\sigma^{\mu\nu}=\frac{i}{4}\left(\sigma^\mu\bar\sigma^\nu-\sigma^\nu\bar\sigma^\mu  \right).
\end{equation}

\section{Determining a basis for the complex four-fermion operators} \label{app:4f_basis} \noindent 
Here we determine all possible four fermion operators respecting the symmetries of the model.  The operators must be singlets under $ \SU(N_\tcf) $, $ \Sp(2N_\tcs) $, and Lorentz transformation, while being symmetric under exchange of the external fermions. The Lorentz contractions are denoted with parenthesis, $ \psi^{i_1} \phantom{}_{a_1} \phantom{}^\alpha \spur{i_2}{a_2 \alpha} =(\spur{i_1}{a_1} \spur{i_2}{a_2}) $. 

We start by noting that the operators must have the general form
\begin{equation}
\mathcal{O}^\TC_i = \dfrac{1}{8\Lambda_\TC^2} (\spur{i_1}{a_1} \spur{i_2}{a_2} ) (\spur{i_3}{a_3} \spur{i_4}{a_4} ) R^{a_1 a_2 a_3 a_4}_{i_1 i_2 i_3 i_4},
\end{equation}
where $ R^{a_1 a_2 a_3 a_4}_{i_1 i_2 i_3 i_4} $ is the tensor structure. This is the only kind of Lorentz structure at lowest order as any Lorentz contraction between $ \sigma_\mu $ matrices can be written as a combination of the trivial tensors $ \delta $ and $ \varepsilon $. The tensor $ R $ must satisfy the symmetries 
\begin{equation}
R^{a_1 a_2 a_3 a_4}_{i_1 i_2 i_3 i_4} = R^{a_2 a_1 a_3 a_4}_{i_2 i_1 i_3 i_4} = R^{a_3 a_4 a_1 a_2}_{ i_3 i_4 i_1 i_2}.
\label{eq:sym_constraints}
\end{equation}
corresponding to the exchange of the external fermions (antisymmetric parts gives vanishing contributions). On the other hand $ R $ cannot be totally symmetric under exchange of all the pairs $ (a_s, i_s) $. Otherwise the operator would vanish due to the fermion identity
\begin{equation}
(f_1 f_2)(f_3 f_4) + (f_1 f_3) (f_2 f_4) + (f_1 f_4) (f_2 f_3) = 0.
\end{equation}
$ R $ should furthermore be an invariant under the global symmetries $ \SU(N_\tcf) $ and $ \Sp(2N_\tcs) $. For this purpose the only nontrivial tensors are the antisymmetric fermion condensate formally transforming as $  {\tiny\yng(1,1)}_{\tcf} $ under $ \SU(N_\tcf) $, though the vacuum breaks the symmetry to $ \Sp(N_\tcf) $, and the antisymmetric invariant $ \epsilon $ of $ \Sp(2N_\tcs) $.   

Thus one finds that $ R^{a_1 a_2 a_3 a_4}_{i_1 i_2 i_3 i_4} $ must be a linear combination of the tensors 
\begin{equation}
\Sigma^{a_{\sigma(1)} a_{\sigma(2)}} \Sigma^{a_{\sigma(3)} a_{\sigma(4)}} \epsilon_{i_{\rho(1)} i_{\rho(2)}} \epsilon_{i_{\rho(3)} i_{\rho(4)}},
\end{equation}
where $ \sigma,\rho $ denotes the different permutations of the integers 1 through 4. Down to a multiplicative factor there are only nine different tensors of this type corresponding to the different ways of arranging the $ a_s $ and $ i_s $ indices into pairs. Constraining the tensors to satisfy the symmetry conditions of eq. \eqref{eq:sym_constraints} we find just five operators that span the space of the 4-fermion operators;
\begin{align}
\mathcal{O}_{4f}^6 &= \dfrac{1}{8 \Lambda_\TC^2} (\spur{i_1}{a_1} \spur{i_2}{a_2} ) (\spur{i_3}{a_3} \spur{i_4}{a_4} )  \Sigma^{a_1 a_2} \Sigma^{a_3 a_4} \epsilon_{i_1 i_2} \epsilon_{i_3 i_4} \, ,\\
\mathcal{O}_{4f}^7 &= \dfrac{1}{8 \Lambda_\TC^2} (\spur{i_1}{a_1} \spur{i_2}{a_2} ) (\spur{i_3}{a_3} \spur{i_4}{a_4} )  \left(\Sigma^{a_1 a_4} \Sigma^{a_2 a_3} - \Sigma^{a_1 a_3} \Sigma^{a_2 a_4}\right) \epsilon_{i_1 i_2} \epsilon_{i_3 i_4} \, ,\\
\mathcal{O}_{4f}^8 &= \dfrac{1}{8 \Lambda_\TC^2} (\spur{i_1}{a_1} \spur{i_2}{a_2} ) (\spur{i_3}{a_3} \spur{i_4}{a_4} )  \Sigma^{a_1 a_2} \Sigma^{a_3 a_4} \left(\epsilon_{i_1 i_4} \epsilon_{i_2 i_3} - \epsilon_{i_1 i_3} \epsilon_{i_2 i_4}\right) , \\
\mathcal{O}_{4f}^9 &= \dfrac{1}{8 \Lambda_\TC^2} (\spur{i_1}{a_1} \spur{i_2}{a_2} ) (\spur{i_3}{a_3} \spur{i_4}{a_4} ) \left( \Sigma^{a_1 a_3} \Sigma^{a_2 a_4} \epsilon_{i_1 i_3} \epsilon_{i_2 i_4} + \Sigma^{a_1 a_4} \Sigma^{a_2 a_3} \epsilon_{i_1 i_4} \epsilon_{i_2 i_3}\right),\\
\mathcal{O}_{4f}^{10} &= \dfrac{1}{8 \Lambda_\TC^2} (\spur{i_1}{a_1} \spur{i_2}{a_2} ) (\spur{i_3}{a_3} \spur{i_4}{a_4} ) \left( \Sigma^{a_1 a_3} \Sigma^{a_2 a_4} \epsilon_{i_1 i_4} \epsilon_{i_2 i_3} + \Sigma^{a_1 a_4} \Sigma^{a_2 a_3} \epsilon_{i_1 i_3} \epsilon_{i_2 i_4} \right).
\end{align}	
The constraint that a tensor which is totally symmetric under exchange of the pairs $ (a_s, i_s) $ leads to a vanishing operators, implies the following linear dependence between the operators:   
\begin{equation}
\mathcal{O}_{4f}^6 + \mathcal{O}_{4f}^9 = \mathcal{O}_{4f}^7 + \mathcal{O}_{4f}^8 - \mathcal{O}_{4f}^{10} = 0.
\end{equation}  
Having used all the constraints on the operators, we find that $ \mathcal{O}_{4f}^6 $, $ \mathcal{O}_{4f}^7 $, and $ \mathcal{O}_{4f}^8 $ make up a basis for the complex 4-fermion operators.

We note that the basis for the self-conjugate 4-fermion operators follows similarly, by noticing that any any Lorentz structure reduces to the forms $ (\spur{i_1}{a_1} \spur{i_2}{a_2}) (\spurbar{i_3}{a_3} \spurbar{i_4}{a_4}) $.

\section{List of NLO kinetic operators} \label{app:pNGB_kinteic}

In this appendix we list the remaining NLO operators for the chiral kinetic term, arising through loop corrections from SM fermions. All these operators contain two derivatives of the pNGB field and some symmetry breaking parameter(s). In the list we have ignored all the terms on the form
	\begin{equation}
	C \Tr \left[(\Sigma^\dagger \overleftrightarrow{D}^\mu \Sigma)^2 \right] \propto \Tr[u_\mu u^\mu]\, , 
	\end{equation}
for some constant $ C $, as these can be reabsorbed into a renormalization of the LO kinetic term. Furthermore we have utilized the fact that  
	\begin{equation} \label{eq:Maurer-Cartan_trace}
	\Tr \left[ (D_\mu \Sigma) \Sigma^\dagger \right] = -i\, \Tr\left[u u_\mu u^\dagger\right] = -i\, \Tr \, u_\mu = 0,
	\end{equation}	 
as the Maurer-Cartan form $ u_\mu $ takes values in the Lie algebra of $ \SU(4)_\tcf $. Any potential term containing this structure has thus been ignored. 

The above consideration leave just one nontrivial, $ \SU(4)_\tcf $ invariant kinetic term with only one insertion of $ y_f^\ast y_f $:
\begin{align}
	\mathcal{O}_{y\Pi D} &= 
	\dfrac{f_\TC^2}{4\pi}\Yf{1}{2}{1}{2} \DSd{1}{3}\DS{3}{2} \epsilon_{i_1 i_2} \, .
\end{align}

With two insertions of $ y_f^\ast y_f $ there are a total of 6 different contractions of the $ \SU(N_\tcf) $ indices and each of these have 3 different ways of contracting the $ \Sp(2N_\tcs) $ indices, only one of which is listed here (the naming is for all three operators). 
These operators are:
\begin{eqnarray}
\mathcal{O}_{y\Pi D}^{1-3} &=& \dfrac{1}{4} \dfrac{\Lambda_\TC^2}{16 \pi^2} \Yf{1}{2}{1}{2} \Yf{3}{4}{3}{4} \SDS{1}{2} \SDS{3}{4} \epsilon_{i_1 i_2} \epsilon_{i_3 i_4}\, ,\\
\mathcal{O}_{y\Pi D}^{4-6} &=& \dfrac{1}{4} \dfrac{\Lambda_\TC^2}{16 \pi^2} \Yf{1}{2}{1}{2} \Yf{3}{4}{3}{4} \SDS{1}{4} \SDS{3}{2}\epsilon_{i_1 i_2} \epsilon_{i_3 i_4}\, , \\
\mathcal{O}_{y\Pi D}^{7-9} &=& \dfrac{\Lambda_\TC^2}{16 \pi^2} \Yf{1}{2}{1}{2} \Yf{3}{4}{3}{4} \DSd{1}{3}\DS{2}{4} \epsilon_{i_1 i_2} \epsilon_{i_3 i_4}\, \\
\mathcal{O}_{y\Pi D}^{10-12} &=& \dfrac{\Lambda_\TC^2}{16 \pi^2}\Yf{1}{1}{1}{2}\Yf{2}{3}{3}{4} \DSd{2}{4}\DS{4}{3}  \epsilon_{i_1 i_2} \epsilon_{i_3 i_4}\, ,\\
\mathcal{O}_{y\Pi D}^{13-15} &=& \dfrac{\Lambda_\TC^2}{16 \pi^2}\Yf{1}{2}{1}{2}\Yf{2}{3}{3}{4}\DSd{1}{4}\DS{4}{3} \epsilon_{i_1 i_2} \epsilon_{i_3 i_4}\, ,\\
\mathcal{O}_{y\Pi D}^{16-18} &=& \dfrac{1}{2} \dfrac{\Lambda_\TC^2}{16 \pi^2} \Yf{1}{2}{1}{2} \Yf{3}{4}{3}{4} \Sigma^\dagger_{a_1a_3} \DS{2}{5} \SDS{5}{4} \epsilon_{i_1 i_2} \epsilon_{i_3 i_4}\, ,
\end{eqnarray}
where the last operator is complex. 

There are 4 real operators with two EW gauge insertion:
\begin{align}
	\mathcal{O}_{\Pi D}^1 &= \dfrac{1}{4} \dfrac{f_\TC^2}{16 \pi^2} \Tr \left[ (\Sigma \overleftrightarrow{D}_\mu \Sigma^\dagger) T_\tcf^{I} (\Sigma \overleftrightarrow{D}^\mu \Sigma^\dagger) T_\tcf^{I}  \right] \, ,\\
	\mathcal{O}_{\Pi D}^2 &= \dfrac{1}{4} \dfrac{f_\TC^2}{16 \pi^2} \Tr \left[ (\Sigma \overleftrightarrow{D}_\mu \Sigma^\dagger) T_\tcf^{I} \right] \Tr \left[ (\Sigma \overleftrightarrow{D}^\mu \Sigma^\dagger) T_\tcf^{I}  \right]\, ,\\
	\mathcal{O}_{\Pi D}^3 &= \dfrac{f_\TC^2}{16 \pi^2} \Tr\left[ (D_\mu \Sigma) (D^\mu \Sigma)^\dagger T_\tcf^{I} T_\tcf^{I} \right]\, ,\\
	\mathcal{O}_{\Pi D}^4 &= \dfrac{f_\TC^2}{16 \pi^2} \Tr\left[ (D_\mu \Sigma) (T_\tcf^{I})\transpose (D^\mu \Sigma)^\dagger T_\tcf^{I} \right]\, 
	\end{align}
where the trace is over the $ \SU(N_\tcf) $ indices.  Additionally  there is 1 complex operator too:
\begin{align}
	\mathcal{O}_{\Pi D}^5 = \dfrac{1}{2} \dfrac{f_\TC^2}{16 \pi^2} \Tr\left[ (D_\mu \Sigma) (\Sigma^\dagger \overleftrightarrow{D}^\mu \Sigma) (T_\tcf^{I})\transpose \Sigma^\dagger T_\tcf^{I} \right]\, .
	\end{align} 
Finally there is one complex term involving the fundamental fermion mass:
	\begin{equation}
	\mathcal{O}_{m\Pi D} = \dfrac{1}{2}\dfrac{f_\TC^2}{16\pi^2} \Tr \left[(D_\mu \Sigma) \chi^\ast (\Sigma \overleftrightarrow{D}^\mu \Sigma^\dagger) \right]\, .
	\end{equation}

\section{List of four-fermion operators} \label{app:list}
We now list all the four-fermion operators found in the model containing only top and bottom SM fermions. These are found by expanding the operators $ \mathcal{O}_{4f}^{1,\ldots,8} $. As it is usually done color indices are always contracted along the spinor structure, and where need we have made use of the $ \SU(3)_c $ generators $ T^{A} = \tfrac{1}{2} \lambda^{A} $. 

Operators with four left-handed quarks: 
\begin{align}
\mathcal{L}_{\mathrm{EFT}} \supset& \quad \dfrac{C_{4f}^{4} + C_{4f}^{5}}{4} \dfrac{|y_{Q_3}|^4}{\Lambda_\TC^2} \left[(\bar{t}_\LL \gamma_\mu t_\LL) (\bar{t}_\LL \gamma^\mu t_\LL) + (\bar{b}_\LL \gamma_\mu b_\LL) (\bar{b}_\LL \gamma^\mu b_\LL)\right] \nonumber \\
& +\dfrac{ c_\theta^2 C_{4f}^{3} +C_{4f}^{4} }{2} \dfrac{|y_{Q_3}|^4}{\Lambda_\TC^2} (\bar{b}_\LL \gamma_\mu b_\LL) (\bar{t}_\LL \gamma^\mu t_\LL) + \dfrac{ -c_\theta^2 C_{4f}^{3} +C_{4f}^{5} }{2} \dfrac{|y_{Q_3}|^4}{\Lambda_\TC^2} (\bar{b}_\LL \gamma_\mu t_\LL) (\bar{t}_\LL \gamma^\mu b_\LL). 
\end{align}
Operators with four right-handed quarks:
\begin{align}
\mathcal{L}_{\mathrm{EFT}} \supset& \quad \dfrac{C_{4f}^{4} + C_{4f}^{5}}{4} \dfrac{|y_t|^4}{\Lambda_\TC^2} (\bar{t}_\RR \gamma_\mu t_\RR) (\bar{t}_\RR \gamma^\mu t_\RR) + \dfrac{C_{4f}^{4} + C_{4f}^{5}}{4} \dfrac{|y_b|^4}{\Lambda_\TC^2}  (\bar{b}_\RR \gamma_\mu b_\RR) (\bar{b}_\RR \gamma^\mu b_\RR) \nonumber \\
& +\dfrac{ c_\theta^2 C_{4f}^{3} +C_{4f}^{4} }{2} \dfrac{|y_t y_b|^2}{\Lambda_\TC^2} (\bar{b}_\RR \gamma_\mu b_\RR) (\bar{t}_\RR \gamma^\mu t_\RR) + \dfrac{ -c_\theta^2 C_{4f}^{3} +C_{4f}^{5} }{2} \dfrac{|y_t y_b|^2}{\Lambda_\TC^2} (\bar{b}_\RR \gamma_\mu t_\RR) (\bar{t}_\RR \gamma^\mu b_\RR). 
\end{align} 
Operators with two left-handed and two right-handed top quarks:
\begin{align}
\mathcal{L}_{\mathrm{EFT}} \supset& \quad \left(-s_\theta^2 C_{4f}^{1} + C_{4f}^{2} \right) \dfrac{|y_{Q_3} y_t|^2}{\Lambda_\TC^2} (\bar{t}_\RR t_\LL) (\bar{t}_\LL t_\RR) - \dfrac{ s_\theta^2 C_{4f}^{3} +C_{4f}^{4} }{2} \dfrac{|y_{Q_3} y_t|^2}{\Lambda_\TC^2}  (\bar{t}_\LL \gamma_\mu t_\LL) (\bar{t}_\RR \gamma^\mu t_\RR) \nonumber \\
& + \left(\dfrac{s_\theta^2 y_{Q_3}^2 y_t^2}{\Lambda_\TC^2} \left[\dfrac{3C_{4f}^6 -3C_{4f}^7 -C_{4f}^8}{6} (\bar{t}_\RR t_\LL) (\bar{t}_\RR t_\LL) - C_{4f}^8  (\bar{t}_\RR T^{A} t_\LL) (\bar{t}_\RR T^{A} t_\LL)  \right] \hc \right). 
\end{align} 	
Operators with two left-handed and two right-handed bottom quarks:
\begin{align}
\mathcal{L}_{\mathrm{EFT}} \supset& \quad \left(-s_\theta^2 C_{4f}^{1} + C_{4f}^{2} \right) \dfrac{|y_{Q_3} y_b|^2}{\Lambda_\TC^2} (\bar{b}_\RR b_\LL) (\bar{b}_\LL b_\RR) - \dfrac{ s_\theta^2 C_{4f}^{3} +C_{4f}^{4} }{2} \dfrac{|y_{Q_3} y_b|^2}{\Lambda_\TC^2}  (\bar{b}_\LL \gamma_\mu b_\LL) (\bar{b}_\RR \gamma^\mu b_\RR) \nonumber \\
& + \left(\dfrac{s_\theta^2 y_{Q_3}^2 y_b^2}{\Lambda_\TC^2} \left[\dfrac{3C_{4f}^6 -3C_{4f}^7 -C_{4f}^8}{6} (\bar{b}_\RR b_\LL) (\bar{b}_\RR b_\LL) - C_{4f}^8  (\bar{b}_\RR T^{A} b_\LL) (\bar{b}_\RR T^{A} b_\LL)  \right] \hc \right). 
\end{align}
Operators with two left-handed and two right-handed quarks, either top and bottom respectively or vice versa:
\begin{align}
\mathcal{L}_{\mathrm{EFT}} \supset& \quad C_{4f}^{2} \dfrac{|y_{Q_3}|^2}{\Lambda_\TC^2} \left[|y_t|^2 (\bar{t}_\RR b_\LL) (\bar{b}_\LL t_\RR) + |y_b|^2 (\bar{t}_\LL b_\RR) (\bar{b}_\RR t_\LL) \right] \nonumber \\
& - \dfrac{C_{4f}^{4} }{2} \dfrac{|y_{Q_3}|^2}{\Lambda_\TC^2} \left[|y_t|^2 (\bar{b}_\LL \gamma_\mu b_\LL) (\bar{t}_\RR \gamma^\mu t_\RR) + |y_b|^2 (\bar{t}_\LL \gamma_\mu t_\LL) (\bar{b}_\RR \gamma^\mu b_\RR)\right]. 
\end{align}
Operators with a left-handed and right-handed top quark, and a left-handed and right-handed bottom quark:
\begin{align}
\mathcal{L}_{\mathrm{EFT}} \supset &-C_{4f}^{1} \dfrac{s_\theta^2 |y_{Q_3}|^2}{\Lambda_\TC^2} \left[y_t y_b^\ast (\bar{t}_\RR t_\LL) (\bar{b}_\LL b_\RR) + y_b y_t^\ast (\bar{b}_\RR b_\LL) (\bar{t}_\LL t_\RR) \right] \nonumber \\
& - \dfrac{C_{4f}^{3}}{2} \dfrac{s_\theta^2 |y_{Q_3}|^2}{\Lambda_\TC^2} \left[y_t y_b^\ast (\bar{b}_\LL \gamma_\mu t_\LL) (\bar{t}_\RR \gamma^\mu b_\RR) + y_b y_t^\ast (\bar{t}_\LL \gamma_\mu b_\LL) (\bar{b}_\RR \gamma^\mu t_\RR) \right] \nonumber \\
&+ \left( 2C_{4f}^8 \dfrac{y_{Q_3}^2 y_t y_b}{\Lambda_\TC^2} \left[ c_{2\theta} (\bar{b}_\RR T^{A} b_\LL) (\bar{t}_\RR T^{A} t_\LL) - c_\theta^2 (\bar{b}_\RR T^{A} t_\LL) (\bar{t}_\RR T^{A} b_\LL) \right] \right. \nonumber \\
&\quad + \dfrac{-3C_{4f}^7 + 2 c_\theta^2 C_{4f}^8}{3} \dfrac{y_{Q_3}^2 y_t y_b}{\Lambda_\TC^2} (\bar{b}_\RR t_\LL) (\bar{t}_\RR b_\LL) \nonumber \\
&\quad +\left. \dfrac{3s_\theta^2 C_{4f}^6 +3 c_\theta^2 C_{4f}^7 - (1+ c_\theta^2) C_{4f}^8}{3} \dfrac{y_{Q_3}^2 y_t y_b}{\Lambda_\TC^2}  (\bar{b}_\RR b_\LL) (\bar{t}_\RR t_\LL) \hc \right).
\end{align}

\bibliography{MPClitt.bib}

\end{document}